\documentclass[%
 reprint,
 amsmath,amssymb,
 aps,
]{revtex4-2}

\usepackage[pdftex]{graphicx}

\graphicspath{
{./Figures/},
{./PhaseDiagrams/},
{./DTSPs/},
}

\usepackage{diagbox}
\usepackage{braket}
\usepackage{dcolumn}
\usepackage{bm}
\usepackage{booktabs}
\usepackage{threeparttable}

\begin{document}

\preprint{APS/123-QED}

\title{Diverse densest ternary sphere packings}

\author{Ryotaro Koshoji}
\email{cosaji@issp.u-tokyo.ac.jp}
\affiliation{Institute for Solid State Physics, The University of Tokyo, Kashiwa 277-8581, Japan}
\author{Taisuke Ozaki}%
 \email{t-ozaki@issp.u-tokyo.ac.jp}
\affiliation{Institute for Solid State Physics, The University of Tokyo, Kashiwa 277-8581, Japan}

\date{\today}

\begin{abstract}

The exploration of the densest structures of multi-sized hard spheres under periodic boundary conditions is a fundamental problem in mathematics and a wide variety of sciences including materials science. We present our exhaustive computational exploration of the densest ternary sphere packings (DTSPs) for 451 radius ratios and 436 compositions on top of our previous study [Koshoji and Ozaki, Phys. Rev. E 104, 024101 (2021)]. The unbiased exploration by a random structure searching method discovers diverse 22 putative DTSPs, and thereby 60 putative DTSPs are identified in total including the 38 DTSPs discussed by the previous study. Some of the discovered DTSPs are well-ordered, for example, the medium spheres in the (9-7-3) structure are placed in a straight line with comprising the unit cell, and the DTSP has the $Pm \bar{3}m$ symmetry if the structural distortion is corrected. At a considerable number of radius ratios, the highest packing fractions are achieved by the phase separations consisting of only the FCC and/or the putative densest binary sphere packings (DBSPs) for all compositions, and the tendency is getting evident as the small and medium spheres are getting larger. The result seems to indicate directly that the local structures in the DBSPs may be denser than those consisting of three kinds of spheres. However, the unit cell of undiscovered DTSPs might only be much larger than in this study due to the complexity of  the ternary local structures. Besides, we found that the phase diagrams may be getting complex as the radii of small and medium spheres are getting smaller. Many competing structures complicate the phase diagrams due to the appearance of many kinds of DTSPs in the phase diagrams, and the competitions may hide unknown DTSPs in a very narrow range of radius ratio. Finally, we discuss the correspondence of the DTSPs with real crystals based on the space group. Our study suggests that the diverse structures of DTSPs can be effectively used as structural prototypes for searching ternary, quaternary, and quinary crystal structures.

\end{abstract}

\maketitle

\section{INTRODUCTION}

The longstanding effort to the analytic identification of the densest unary sphere packing (DUSP) has been fulfilled in the 2000s despite the simpleness of the problem~\cite{10.2307/20159940}. Likewise, analytic identification of the densest binary sphere packings (DBSPs) is also too difficult to prove, however, the development of computers has been enabling us to explore the DBSPs by computer simulations~\cite{doi:10.1021/jp804953r, PhysRevE.79.046714, doi:10.1021/jp206115p, doi:10.1021/jp1045639, Hudson_2011, doi:10.1063/1.5052478, de_laat_de_oliveira_filho_vallentin_2014, PhysRevLett.107.125501, PhysRevE.85.021130}. As a result, a total of 28 putative DBSPs are known at the present time~\cite{PhysRevE.103.023307}.

If we only explore the periodic dense packings of spheres of $n$ different sizes by computer, the packing fraction at given sphere-composition ratio is maximized by filling the space with not more than $n$ kinds of periodic packings~\cite{PhysRevLett.107.125501, PhysRevE.85.021130, PhysRevE.103.023307}. The necessity of the phase separation can be understood from a simple example: At a given composition, the periodic boundary condition limits the maximum packing fraction, and in many cases, it is less than the packing fraction of the FCC packing. In this case, the space-filling with the single periodic packing is sparser than the phase separation to the $n$ kinds of FCC packings. Furthermore, if we know denser multiary sphere packings than the FCC packing, the phase separations including the dense packings have a higher packing fraction than that consisting of only $n$ kinds of FCC packings. The examples indicate that we must construct the phase diagrams, which show the densest phase separation at each radius and composition ratio, so that we identify the densest multiary sphere packings.

\begin{figure*}
\centering
\includegraphics[width=2\columnwidth]{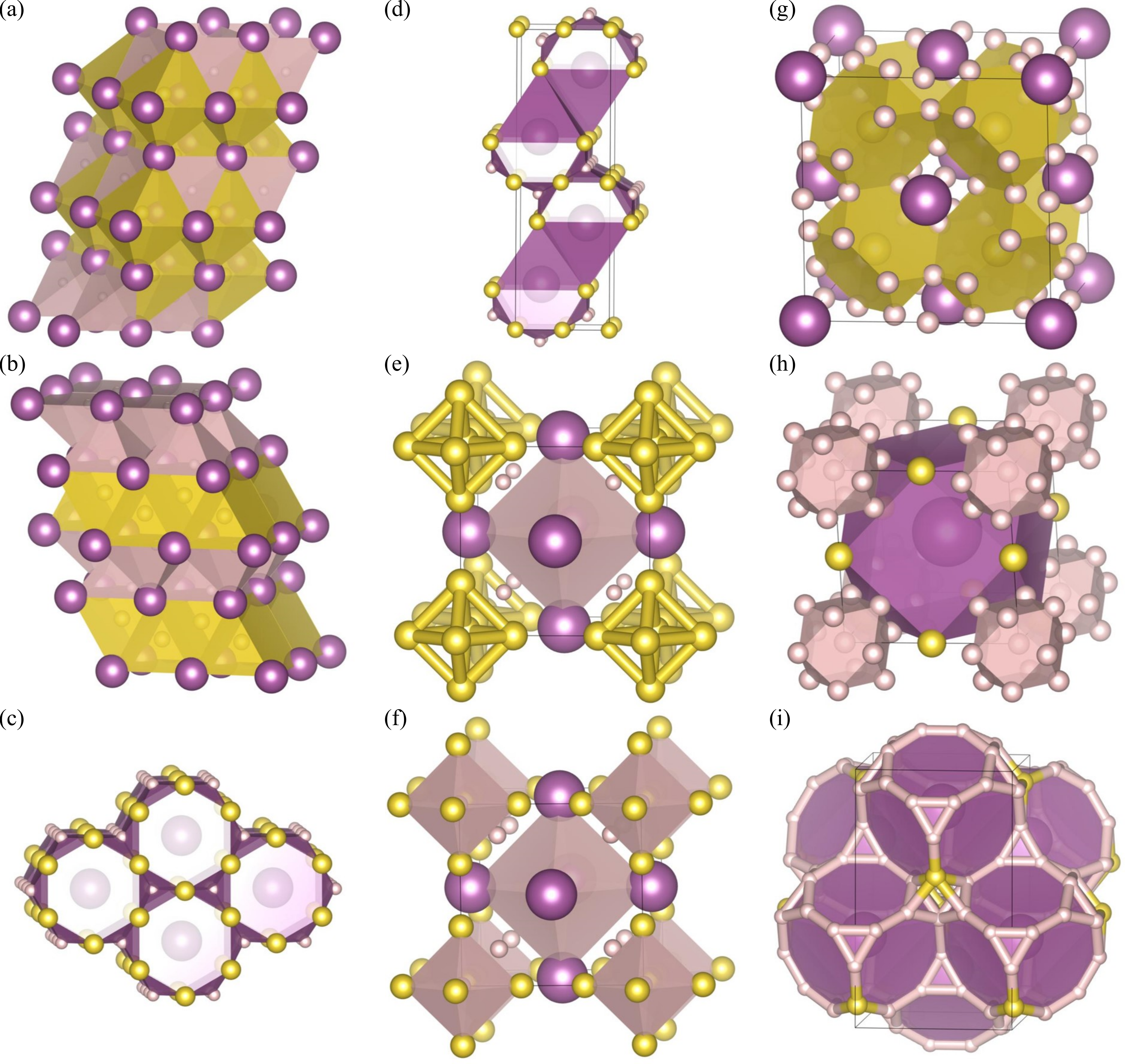}
\caption{The well-ordered putative DTSPs that have already been discovered in Ref.~\cite{PhysRevE.104.024101}. Structural figures in this study are generated by VESTA~\cite{Momma:db5098}. For visibility, some of them are shown in conventional cells with symmetrization. In all figures of the paper, the small, medium, and large spheres are represented by pink, yellow, and purple balls, respectively. (a) One of the (2-2-2) structures, called (2-2-2)$\mathrm{_1}$ structure in this paper. (b) One of the (2-2-2) structures, called (2-2-2)$\mathrm{_2}$ structure in this paper. (c) The (4-3-1) structure. (d) The (4-4-2) structure. (e) The (9-6-3) structure. (f) The (10-6-3) structure. (g) The (13-2-1) structure. (h) The (13-3-1) structure. (i) The (16-2-2) structure.}
\label{fig:knownDTSPs}
\end{figure*}

However, previous studies on the dense ternary sphere packings~\cite{WONG2014357, doi:10.1021/ie200765h, YI2012129, ROQUIER2019343, doi:10.1063/1.4941262} do not construct the phase diagrams. Note that the DBSPs are the candidates for the phase separation as well as the FCC packings of small, medium, and large spheres, so we have to know the precise packing fractions of the DBSPs to construct the phase diagrams of the densenst ternary sphere packings (DTSPs). Accordingly, following the seminal studies which had constructed the phase diagram for binary system~\cite{PhysRevLett.107.125501, PhysRevE.85.021130, PhysRevE.103.023307}, we constructed the phase diagram for ternary systems at 45 kinds of radius ratios so as to find the DTSPs~\cite{PhysRevE.104.024101}. As a result, we successfully discovered the 37 putative DTSPs and also found that the well-ordered DTSPs as exemplified in Fig.~\ref{fig:knownDTSPs} tend to appear on the phase diagrams at radius ratios where the radius ratios of small and medium spheres are relatively large.

As shown in the previous studies~\cite{PhysRevLett.107.125501, PhysRevE.85.021130, PhysRevE.103.023307}, some of the DBSPs such as the (7-3) structure become the densest in very narrow radius ratios, and this is because several packings are competitive to each other with respect to packing fractions at every radius ratio. This complexity of the binary phase diagram indicates that some of the unknown DTSPs are also the densest in very narrow radius ratios. Therefore, it is necessary to explore the DTSPs more precisely with respect to the radius ratios than the previous study~\cite{PhysRevE.104.024101} so that we know the full picture of the ternary phase diagram.

The well-ordered DTSPs may lead to promising structural prototypes for exploring materials under high pressure. In fact, the previous study~\cite{PhysRevE.103.023307} showed that many crystals can be understood as DBSPs. For example, the clathrate crystal of the $\mathrm{LaH}_{10}$, which is one of the superhydrides synthesized under high pressure, corresponds to the $\mathrm{XY}_{10}$ structure that is one of the DBSPs. The hydrogen sublattice of the $\mathrm{LaH}_{10}$ is responsible for high-temperature superconductivity~\cite{PhysRevLett.21.1748, PhysRevLett.78.118, PhysRevLett.92.187002}, and the structural property can be realized by some of the DTSPs such as the (13-2-1) structure~\cite{PhysRevE.104.024101} with replacing small spheres with hydrogen atoms. Although the previous study~\cite{PhysRevE.104.024101} reported that the correspondence of the DTSPs with crystals seems to be exceptional, we can expect that the unique structural prototypes, which is difficult to design based on the perspectives of chemical bonds and/or local polyhedrons as found in coordination structures to transition elements, may enable materials to have functional properties under high pressure.

In this study, to show the full picture of the ternary phase diagram with discovering unknown DTSPs following the previous study~\cite{PhysRevE.104.024101}, we have additionally explored the DTSPs at 451 kinds of radius ratios and 436 kinds of compositions. The maximum number of spheres per unit cell is set to be 25 for most radius ratios, where the number is larger than that in the previous study~\cite{PhysRevE.104.024101}. As a result, we have additionally found 22 putative DTSPs including six structures identified as Semi-DTSPs (SDTSPs) in Ref.~\cite{PhysRevE.104.024101} and the (8-6-2)$_{\mathrm{L}}$ structure that is the long-period structure of the (4-3-1) structure~\cite{PhysRevE.104.024101}. Since the minimum radius ratio of small spheres is set to be larger than in the previous study~\cite{PhysRevE.104.024101}, the discovered DTSPs are well-ordered; for example, the (9-7-3) structure is comprised of the cubic unit cell constituted by medium spheres, and the DTSP has the $Pm \bar{3}m$ symmetry if the structural distortion is corrected. In addition, our results show that at a considerable number of radius ratios, the highest packing fraction at each composition is achieved by the phase separations consisting of only the FCC and/or the DBSPs, and the tendency is getting evident as the small and medium spheres are getting larger. On the other hand, in the case that $\alpha_1 = 0.29$, we find nine DTSPs; the discovery indicates that the phase diagrams may be getting complex as the radii of small and medium spheres are getting smaller.

The paper is organized as follows: Section \ref{sec:method} describes the conditions for the exhautive exploration of the DTSPs; Sec.~\ref{sec:results_and_discussion} presents the discovered DTSPs with phase diagrams and discusses the results. In Sec.~\ref{sec:conclusion}, we summarize this study.

\section{Computational methods and conditions}
\label{sec:method}

\begin{figure}
\centering
\includegraphics[width=0.65\columnwidth]{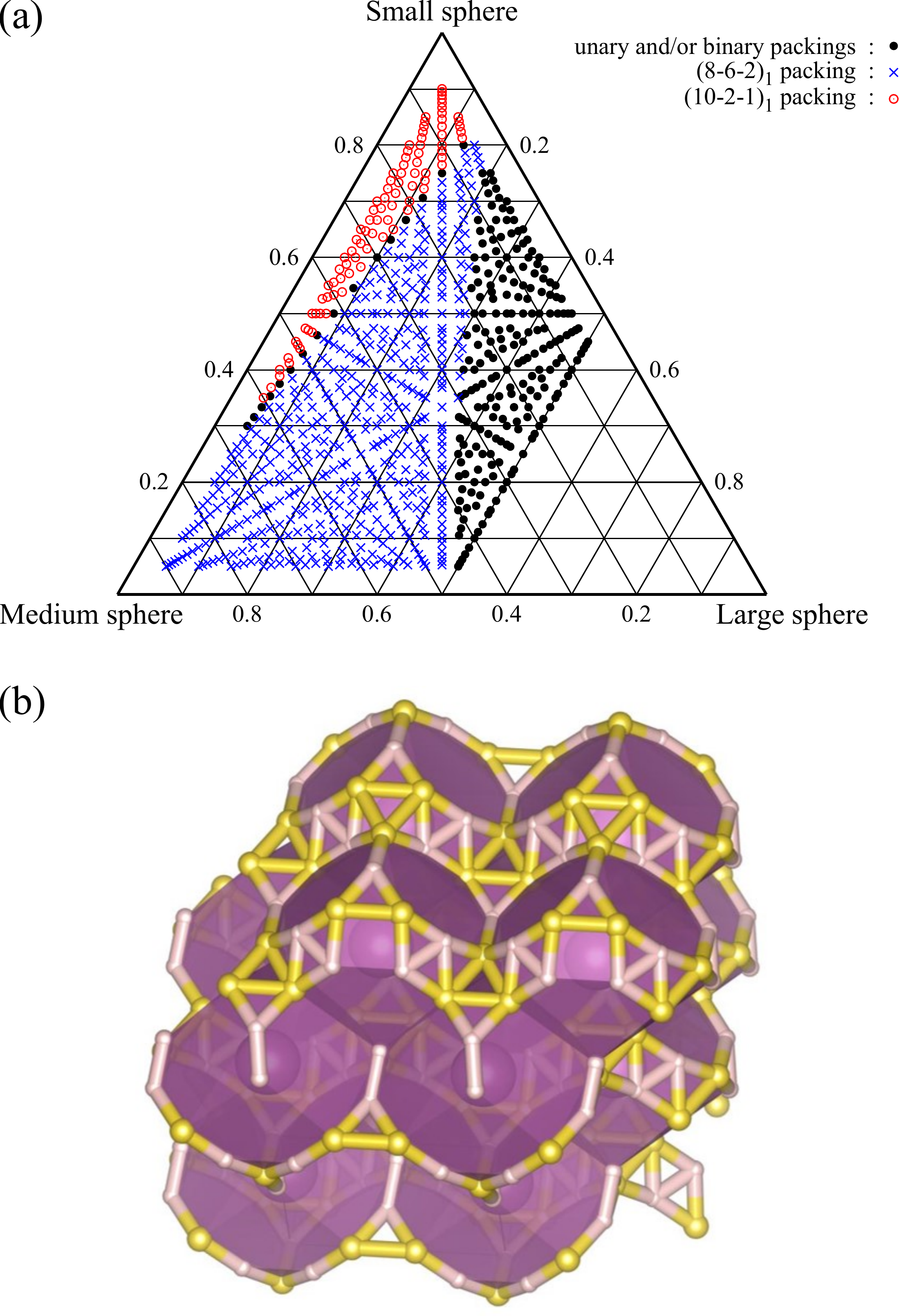}
\caption{The phase diagram at the radius ratio of $0.29:0.39:1.00$ (a) shows that the (8-6-2)$_1$ structure (b) is the DTSP, and the packing fraction is $0.788521$. The (10-2-1)$\mathrm{_1}$ structure plotted on the phase diagram (a) has the packing fraction of $0.793035$, and the structure is shown in Fig.~\ref{fig:029-040-100}(b). Note that the way to read the phase diagrams is discussed in Fig.~2 of Ref.~\cite{PhysRevE.104.024101}, and the lines between spheres in all the structural figures in this study do not necessarily correspond to the contact between spheres.}
\label{fig:029-039-100}
\end{figure}
\begin{figure}
\centering
\includegraphics[width=0.65\columnwidth]{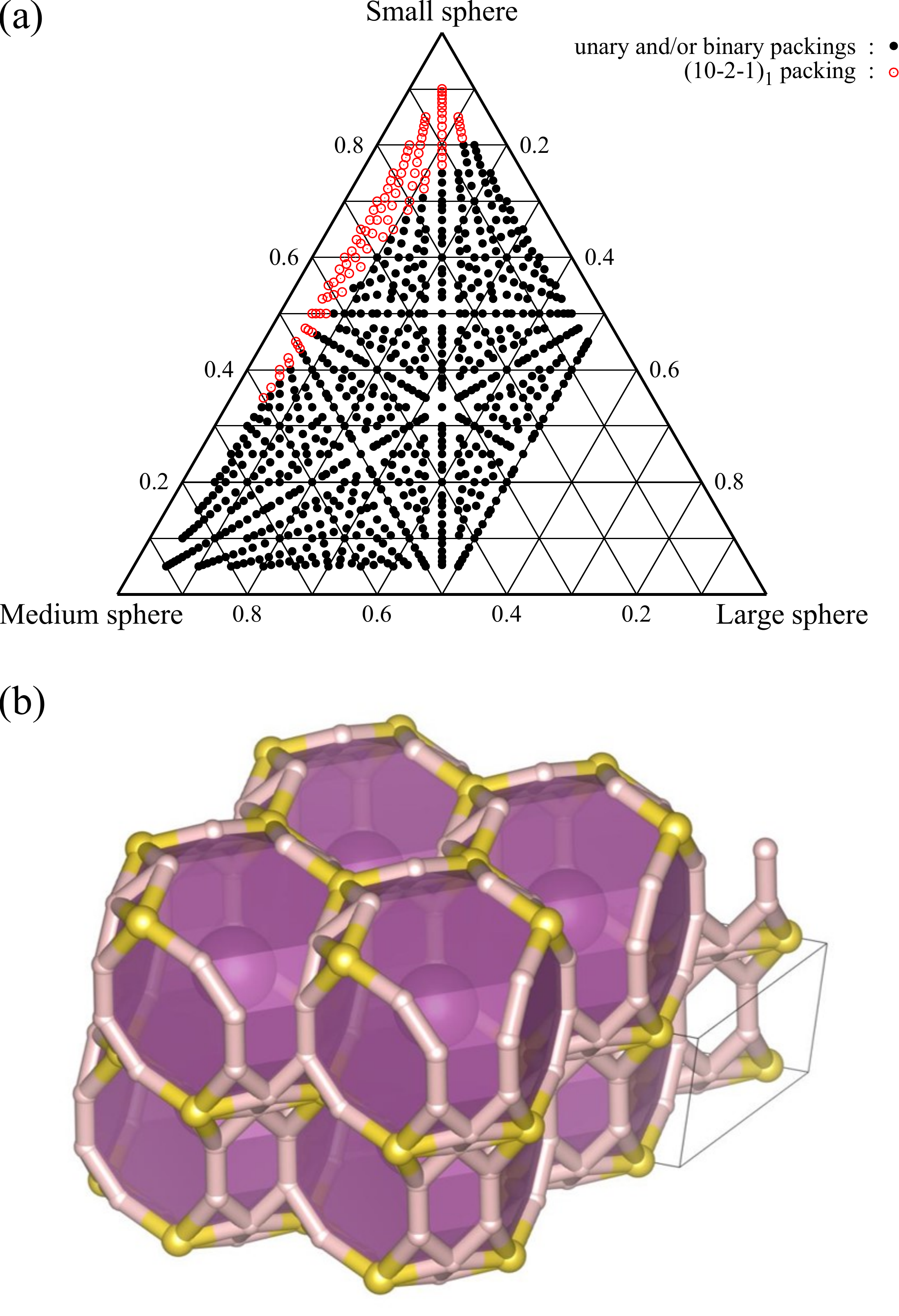}
\caption{The phase diagram at the radius ratio of $0.29:0.40:1.00$ (a) shows that the (10-2-1)$\mathrm{_1}$ structure (b) is the DTSP, and the packing fraction is $0.796814$. This structure also appears on the phase diagram at $0.29:0.39:1.00$ as shown in Fig.~\ref{fig:029-039-100}(a).}
\label{fig:029-040-100}
\end{figure}
\begin{figure}
\centering
\includegraphics[width=0.65\columnwidth]{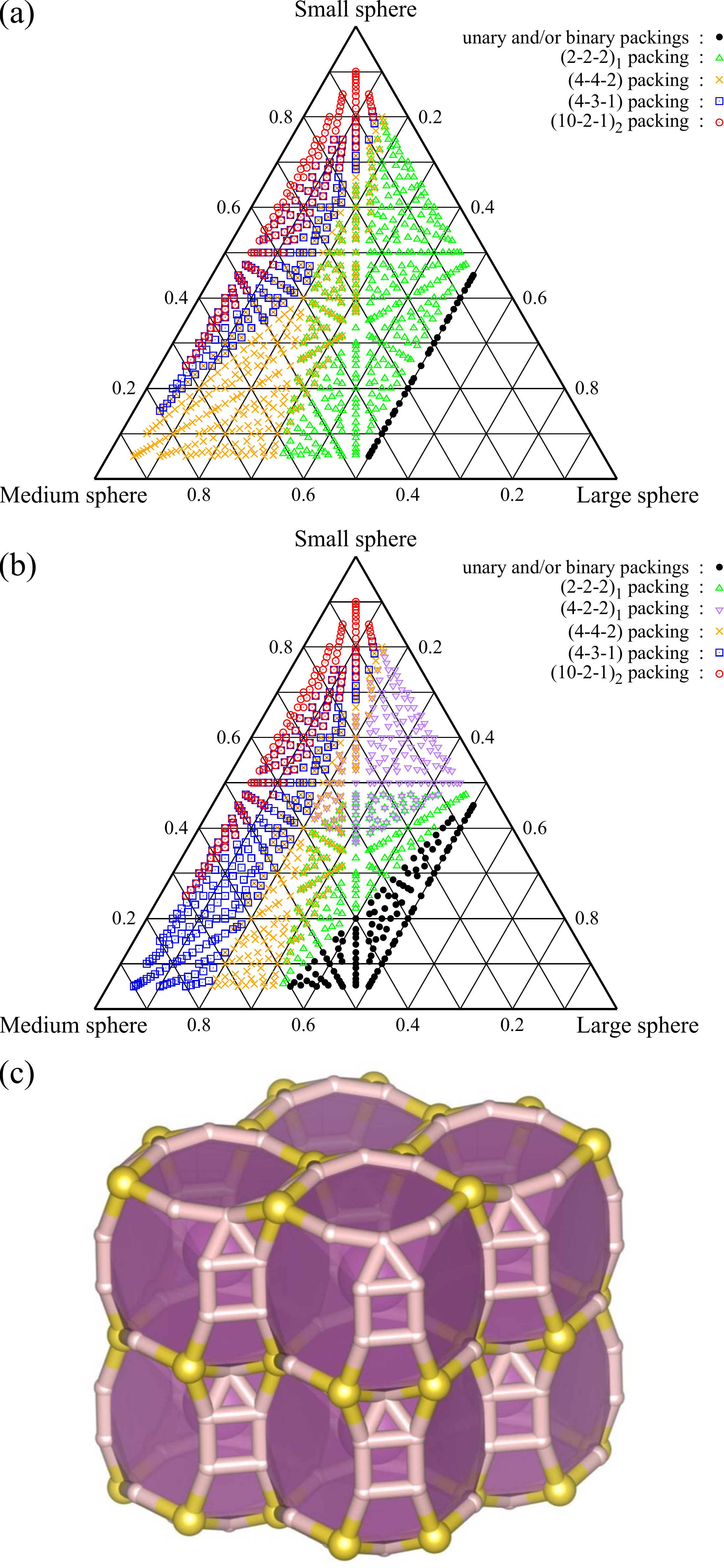}
\caption{The phase diagrams at the radius ratios of $0.29:0.45:1.00$ (a) and $0.29:0.46:1.00$ (b) show that the (10-2-1)$\mathrm{_2}$ structure (c) is the DTSP, and the packing fractions at the two radius ratios are $0.789643$ and $0.790202$, respectively. The (4-2-2)$_1$ structure plotted on the phase diagram (b) has the packing fraction of $0.773744$, and the structure is shown in Fig.~\ref{fig:031-048-100}(d). The (2-2-2)$_1$, (4-3-1), and (4-4-2) structures plotted on the phase diagrams are also the DTSPs shown in Figs.~\ref{fig:knownDTSPs}(a), ~\ref{fig:knownDTSPs}(c), and ~\ref{fig:knownDTSPs}(d), respectively.}
\label{fig:029-045046-100}
\end{figure}
\begin{figure}
\centering
\includegraphics[width=0.65\columnwidth]{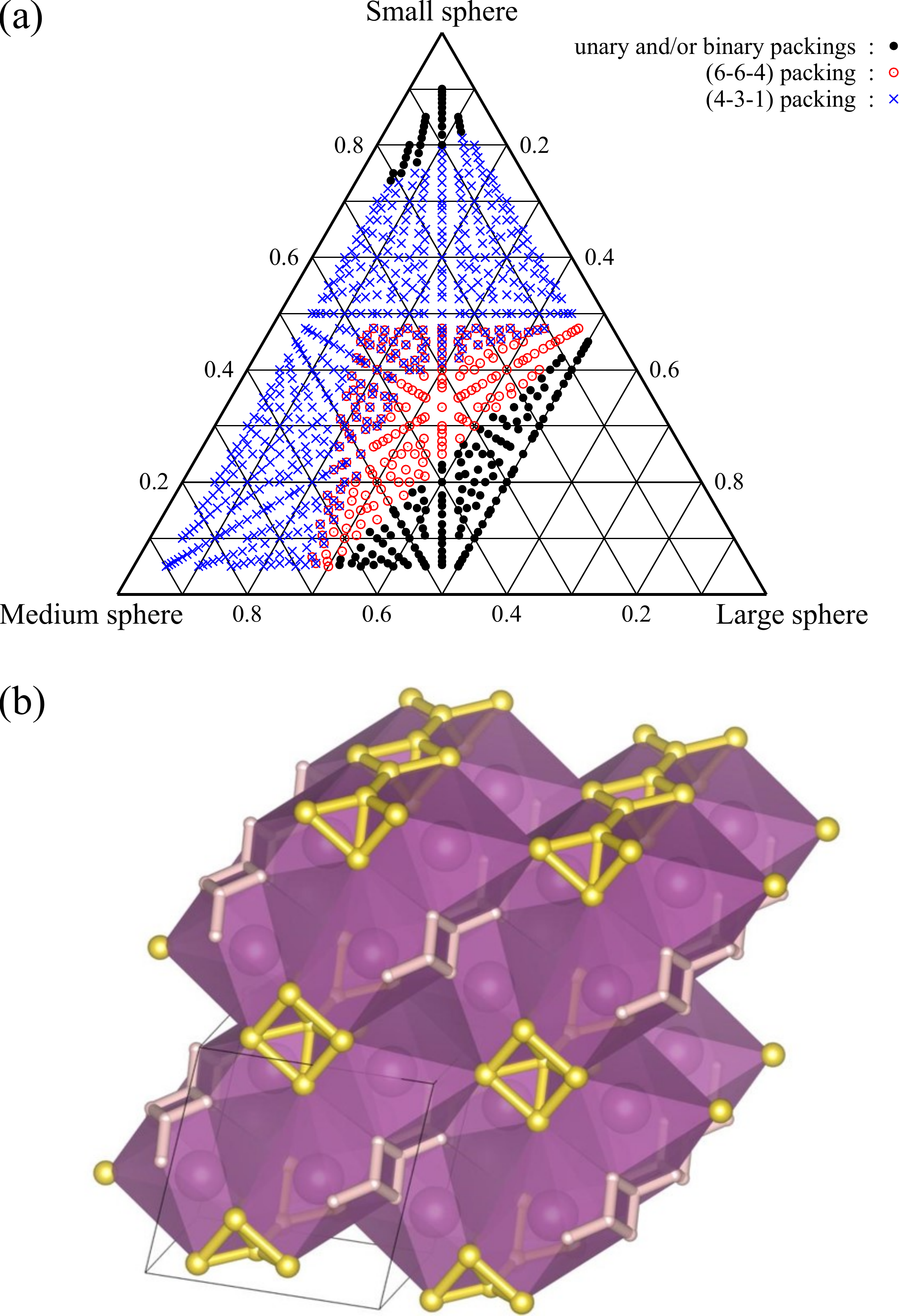}
\caption{The phase diagram at the radius ratio of $0.29:0.48:1.00$ (a) shows that the (6-6-4) structure (b) is the DTSP, and the packing fraction is $0.765610$. If the structural distortion is corrected, this structure has the $I4/m$ symmetry. The (4-3-1) structure plotted on the phase diagram is the DTSP shown in Fig.~\ref{fig:knownDTSPs}(c).}
\label{fig:029-048-100}
\end{figure}
\begin{figure}
\centering
\includegraphics[width=0.65\columnwidth]{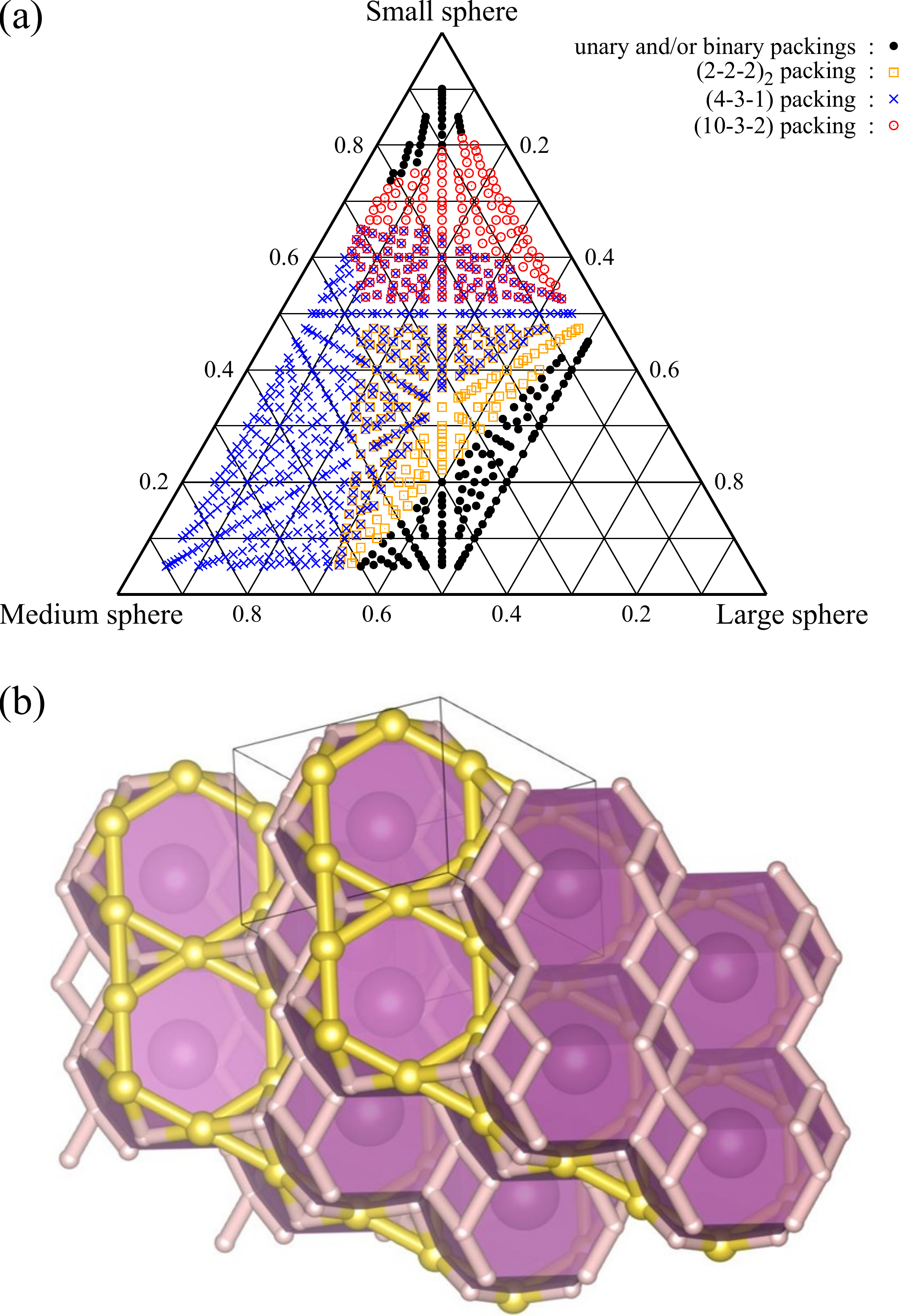}
\caption{The phase diagram at the radius ratio of $0.29:0.50:1.00$ (a) shows that the (10-3-2) structure (b) is the DTSP, and the packing fraction is $0.798053$. This structure, which has already been discussed in the previous study~\cite{PhysRevE.104.024101} as SDTSP, has the $Fmmm$ symmetry if the structural distortion is corrected. The (2-2-2)$_2$ and (4-3-1) structures plotted on the phase diagram are the DTSPs shown in Figs.~\ref{fig:knownDTSPs}(b) and \ref{fig:knownDTSPs}(c), respectively.}
\label{fig:029-050-100}
\end{figure}
\begin{figure}
\centering
\includegraphics[width=0.65\columnwidth]{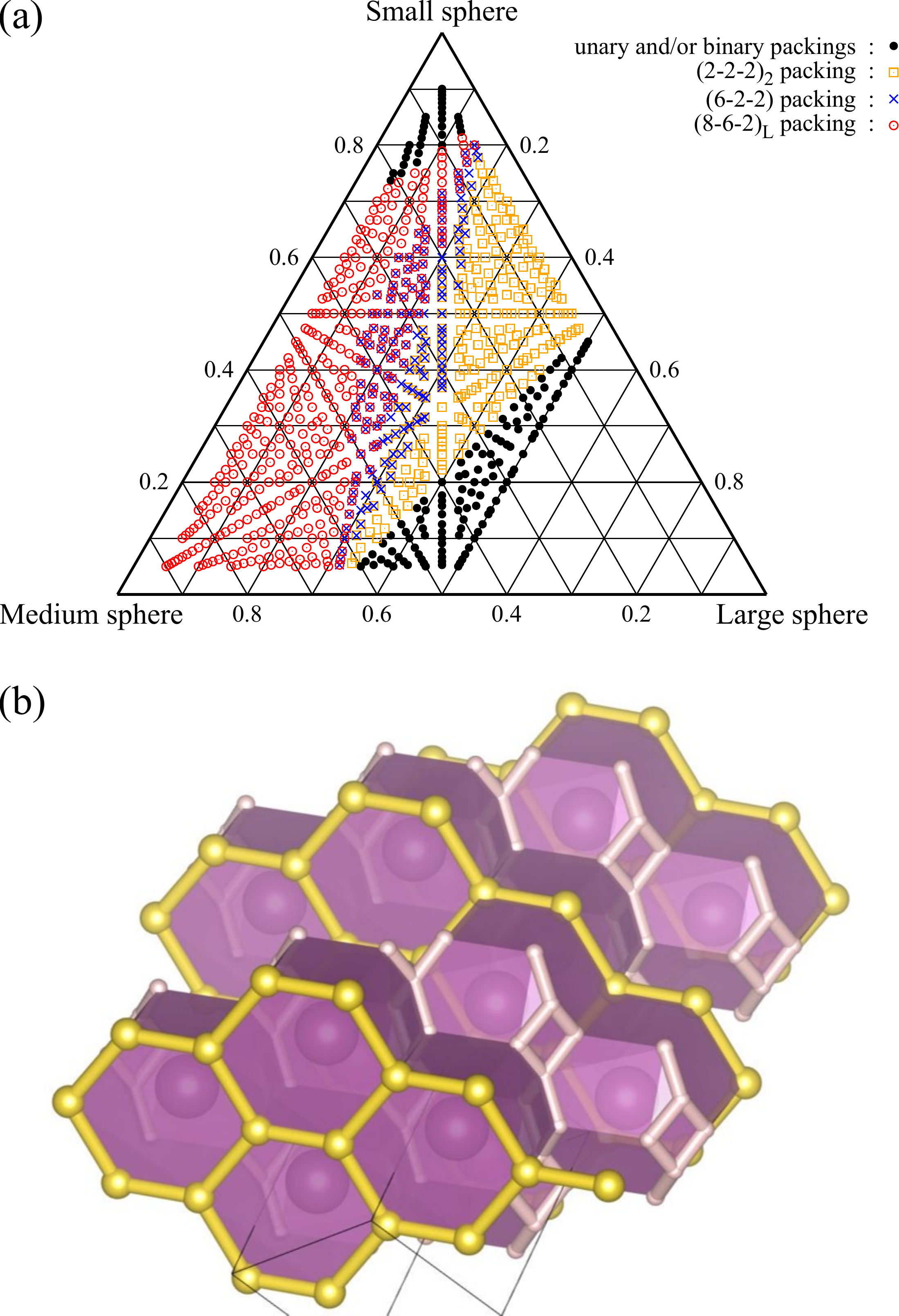}
\caption{The phase diagram at the radius ratio of $0.29:0.51:1.00$ (a) shows that the (6-2-2) (b) structure is the DTSP, and the packing fraction is $0.781880$. This structure has already been discussed in the previous study~\cite{PhysRevE.104.024101} as SDTSP.  The (8-6-2)$_{\mathrm{L}}$ structure plotted on the phase diagram (a) has the packing fraction of $0.780879$, and the structure is shown in Fig.~\ref{fig:030-051-100}(b). The (2-2-2)$_2$ structure plotted on the phase diagram is also the DTSP shown in Figs.~\ref{fig:knownDTSPs}(b).}
\label{fig:029-051-100}
\end{figure}
\begin{figure}
\centering
\includegraphics[width=0.65\columnwidth]{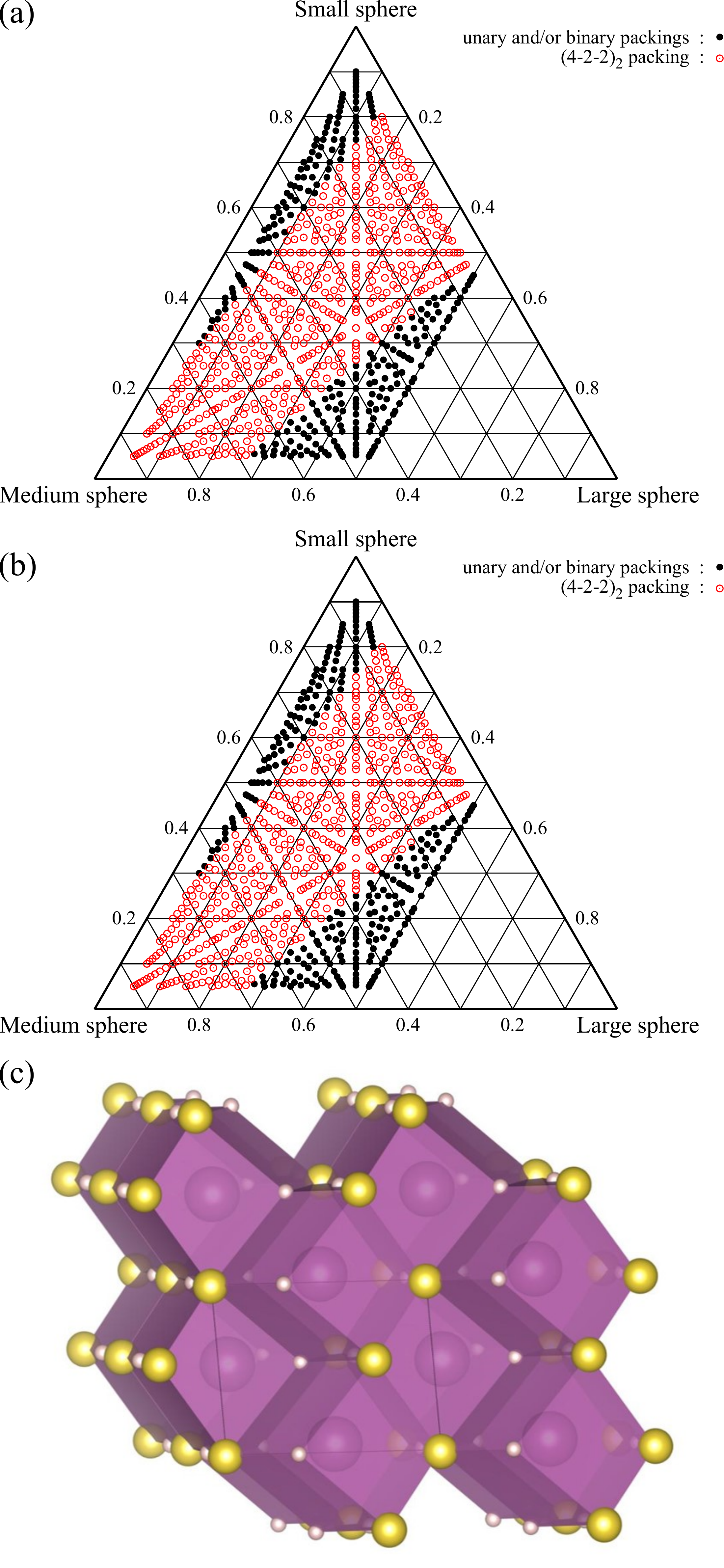}
\caption{The phase diagrams at the radius ratios of $0.29:0.63:1.00$ (a) and $0.29:0.64:1.00$ (b) show that the (4-2-2)$_2$ structure (c) is the DTSP, and the packing fractions at the two radius ratios are $0.764640$ and $0.762837$, respectively. This structure has already been discussed in the previous study~\cite{PhysRevE.104.024101} as SDTSP.}
\label{fig:029-063064-100}
\end{figure}
\begin{figure}
\centering
\includegraphics[width=0.65\columnwidth]{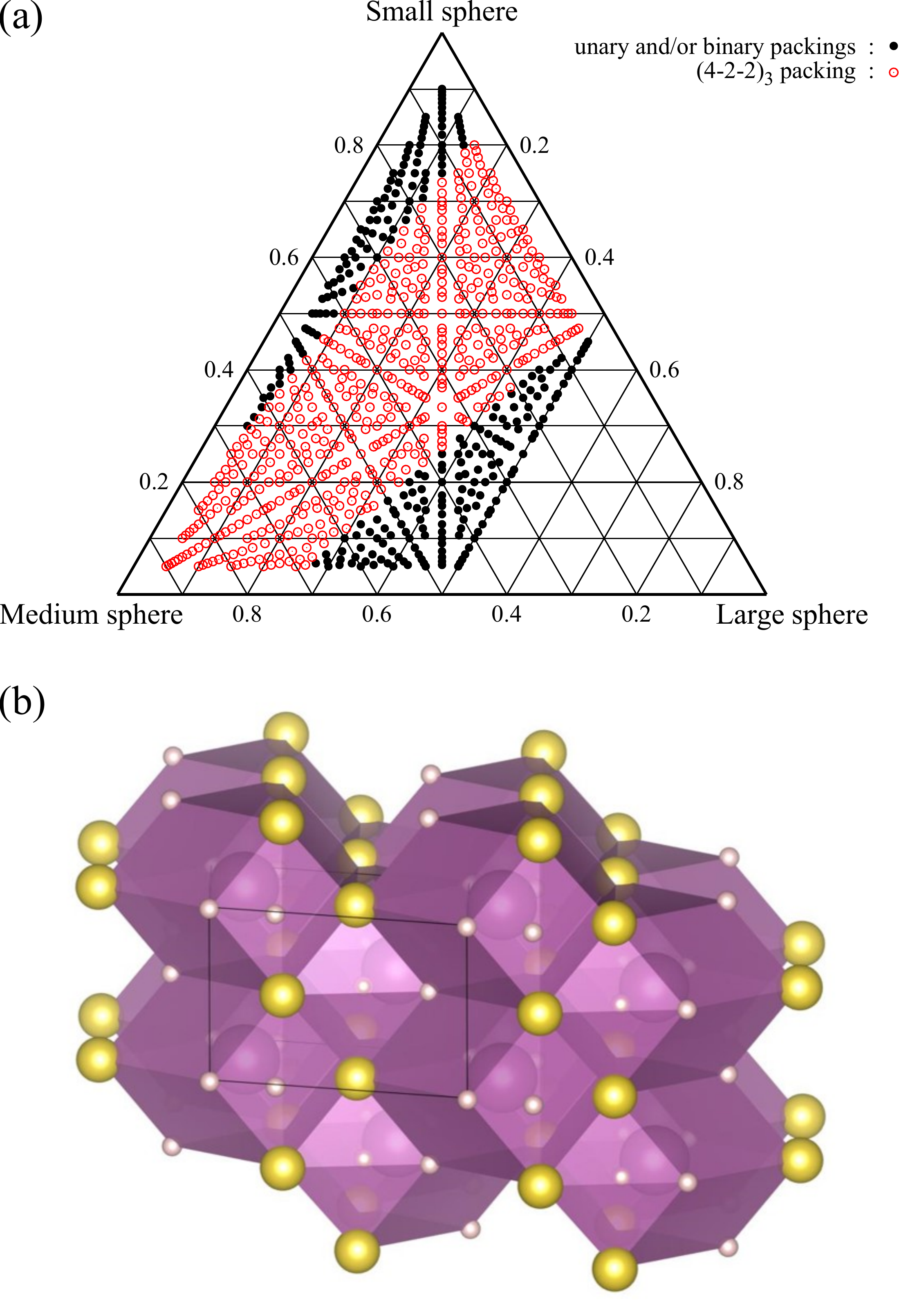}
\caption{The phase diagram at the radius ratio of $0.29:0.65:1.00$ (a) shows that the (4-2-2)$_3$ structure (b) is the DTSP, and the packing fraction is $0.764129$.}
\label{fig:029-065-100}
\end{figure}
\begin{figure}
\centering
\includegraphics[width=0.65\columnwidth]{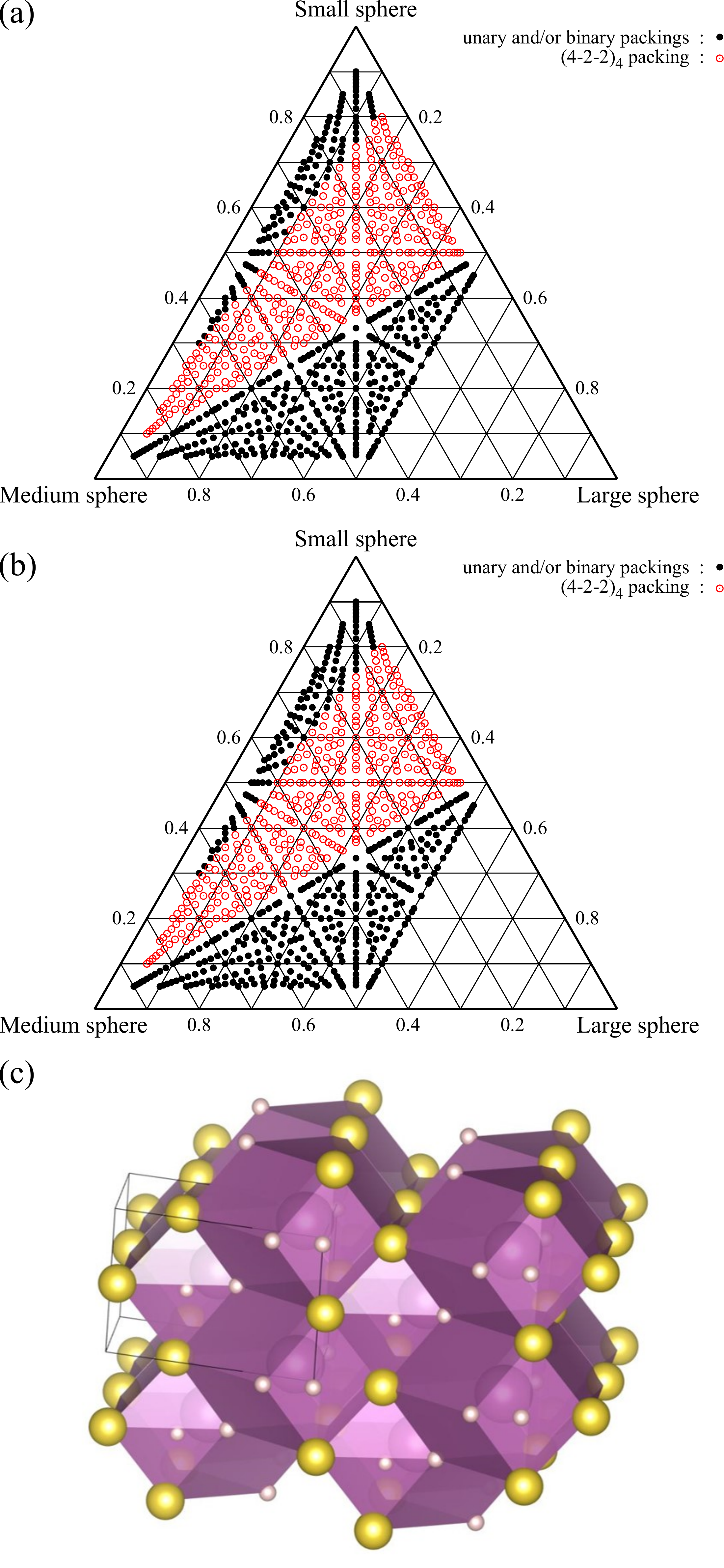}
\caption{The phase diagrams at the radius ratios of $0.29:0.66:1.00$ (a) and $0.30:0.66:1.00$ (b) show that the (4-2-2)$_4$ structure (c) is the DTSP, and the packing fractions at the two radius ratios are $0.761757$ and $0.762522$, respectively.}
\label{fig:029030-066-100}
\end{figure}
\begin{figure*}
\centering
\includegraphics[width=1.6\columnwidth]{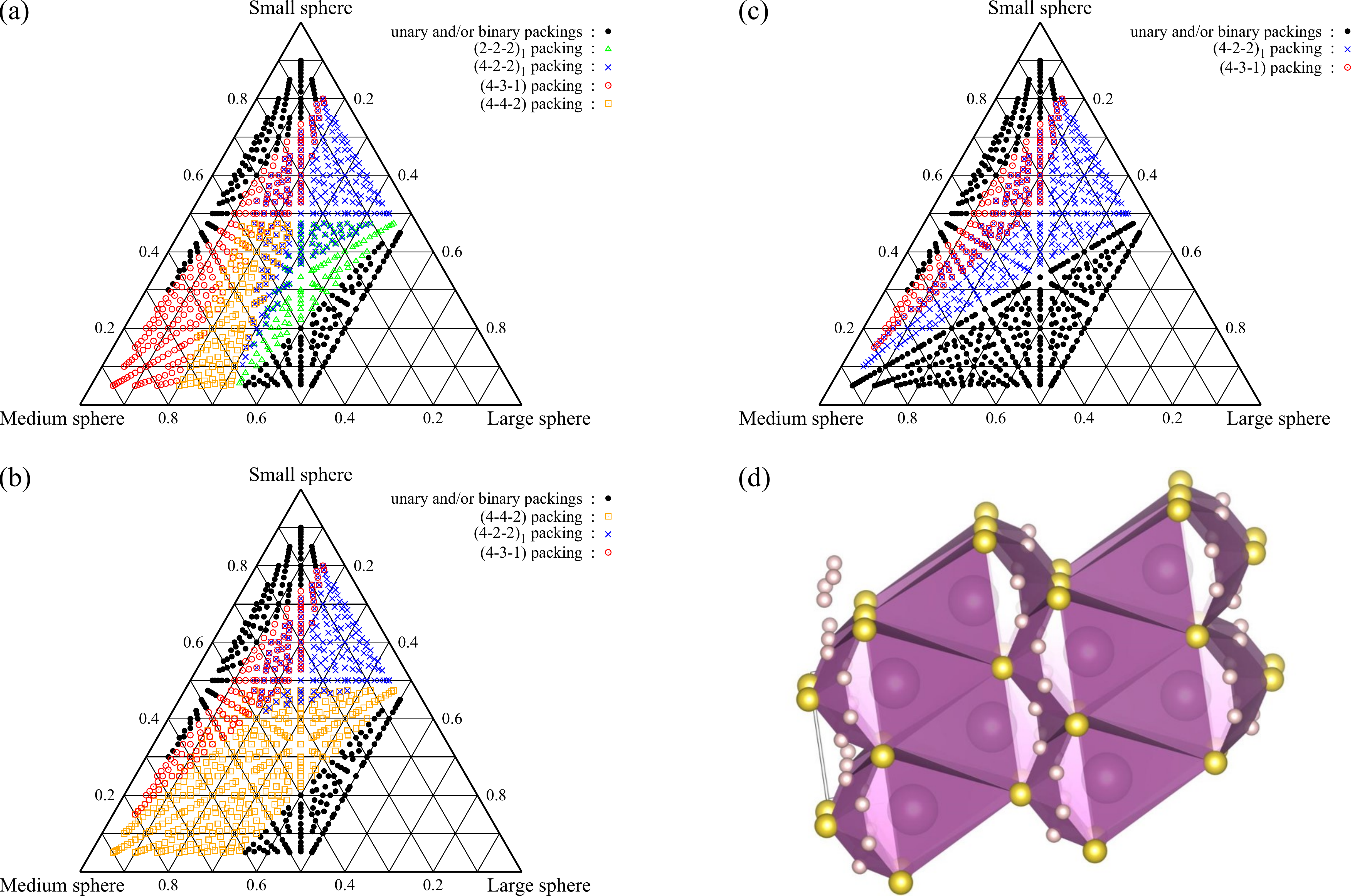}
\caption{The phase diagrams at the radius ratios of $0.30:0.46:1.00$ (a), $0.30:0.47:1.00$ (b) and $0.31:0.48:1.00$ (c) show that the (4-2-2)$\mathrm{_1}$ structure (d) is the DTSP, and the packing fractions at the three radius ratios are $0.773388$, $0.771288$, and $0.768981$, respectively. This structure also appears on the phase diagram at $0.29:0.46:1.00$ as shown in Fig.~\ref{fig:029-045046-100}(b). The (2-2-2)$_1$, (4-3-1), and (4-4-2) structures plotted on the phase diagrams are the DTSPs shown in Figs.~\ref{fig:knownDTSPs}(a), ~\ref{fig:knownDTSPs}(c), and ~\ref{fig:knownDTSPs}(d), respectively.}
\label{fig:031-048-100}
\end{figure*}
\begin{figure}
\centering
\includegraphics[width=0.65\columnwidth]{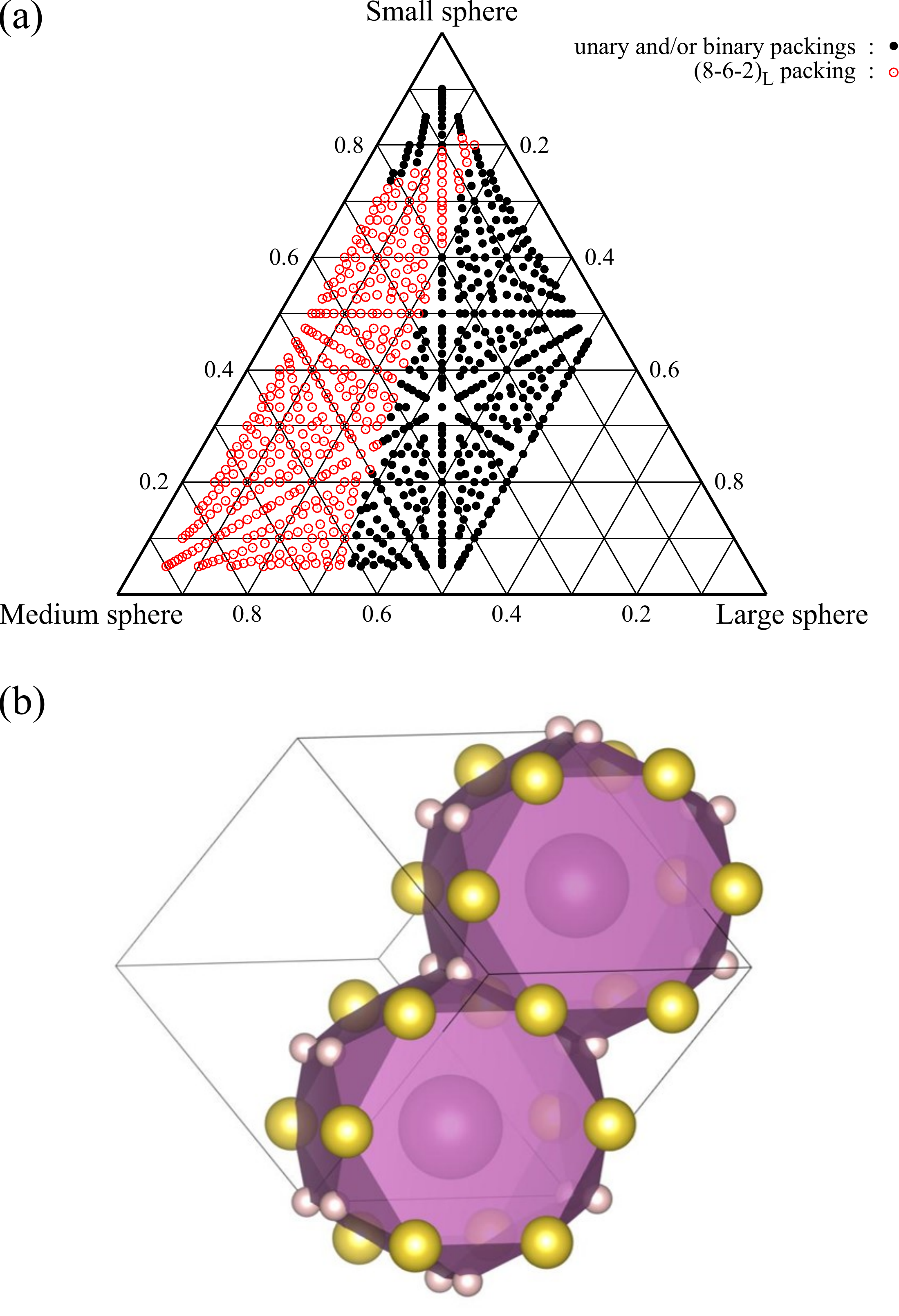}
\caption{The phase diagram at the radius ratio of $0.30:0.51:1.00$ (a) shows that the (8-6-2)$_\mathrm{L}$ structure (b) is the DTSP, and the packing fraction of this structure is $0.786007$. The unit cell of this structure is twice as large as that of the (4-3-1) structure. Note that the packing fractions of the (4-3-1) structure at the two radius ratios of $0.29:0.51:1.00$ and $0.30:0.51:1.00$ are $0.780546$ and $0.785997$, respectively.}
\label{fig:030-051-100}
\end{figure}
\begin{figure}
\centering
\includegraphics[width=0.65\columnwidth]{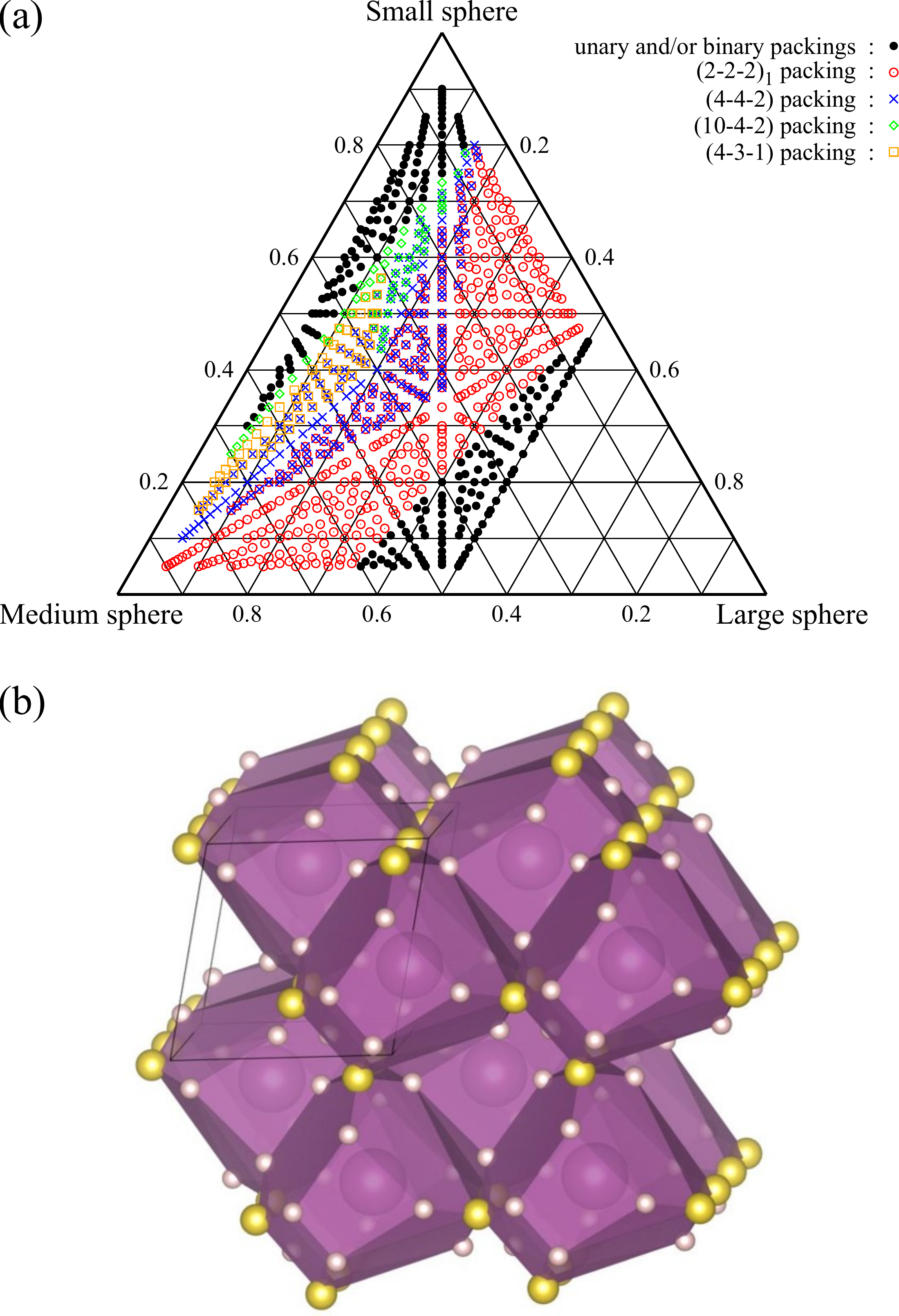}
\caption{The phase diagram at the radius ratio of $0.31:0.47:1.00$ (a) shows that the (10-4-2) structure (b) is the DTSP, and the packing fraction is $0.783260$. If the structural distortion is corrected, this structure has the $Cmcm$ symmetry. The (2-2-2)$_1$, (4-3-1), and (4-4-2) structures plotted on the phase diagram are the DTSPs shown in Figs.~\ref{fig:knownDTSPs}(a), \ref{fig:knownDTSPs}(c), and \ref{fig:knownDTSPs}(d), respectively.}
\label{fig:031-047-100}
\end{figure}
\begin{figure}
\centering
\includegraphics[width=0.65\columnwidth]{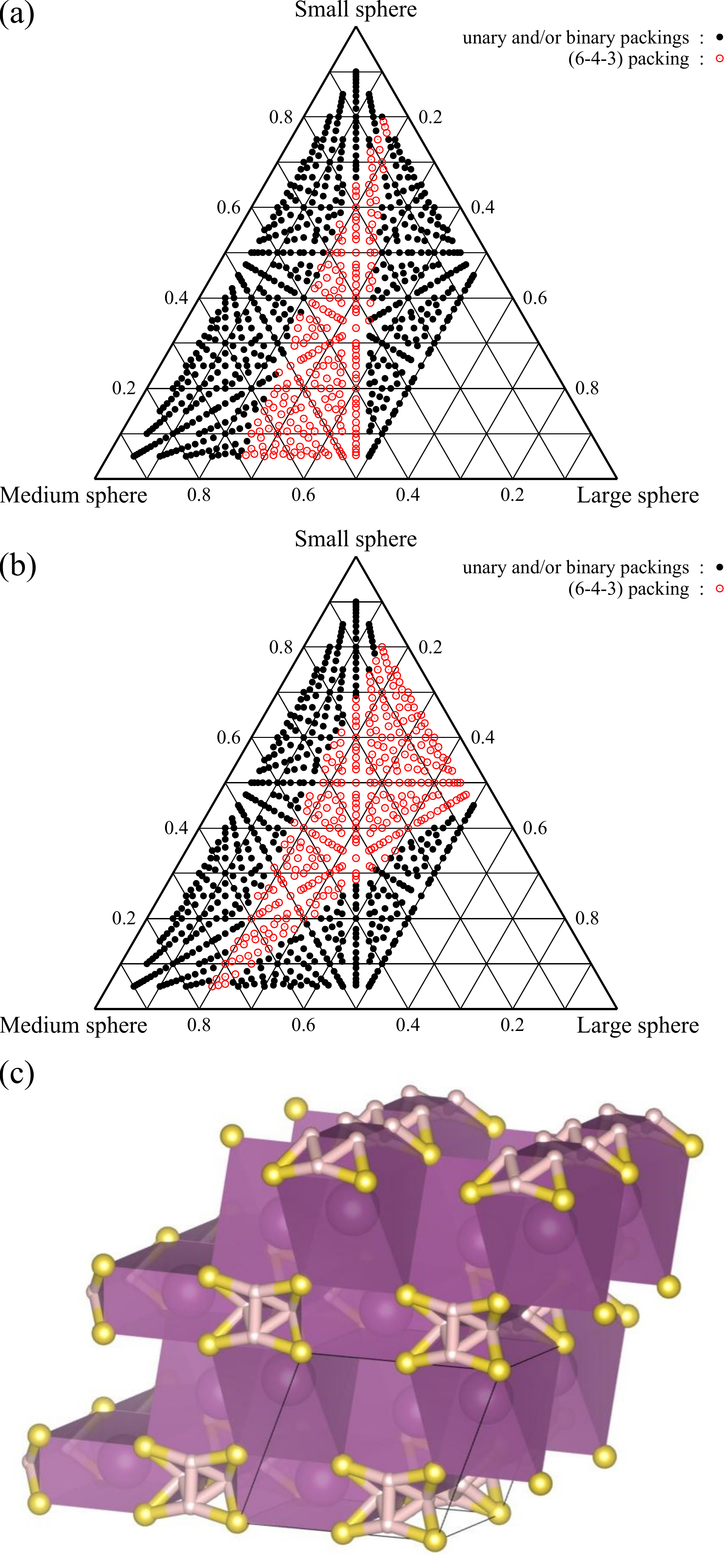}
\caption{The phase diagrams at the radius ratio of $0.33:0.44:1.00$ (a) and $0.33:0.45:1.00$ (b) show that the (6-4-3) structure (c) is the DTSP, and the packing fractions at the two radius ratios are $0.771081$ and $0.768801$, respectively. If the structural distortion is corrected, this structure has the $Immm$ symmetry. This structure also appears on the phase diagram at $0.34:0.45:1.00$ as shown in Fig.~\ref{fig:034-045-100}(a).}
\label{fig:033-044045-100}
\end{figure}
\begin{figure}
\centering
\includegraphics[width=0.65\columnwidth]{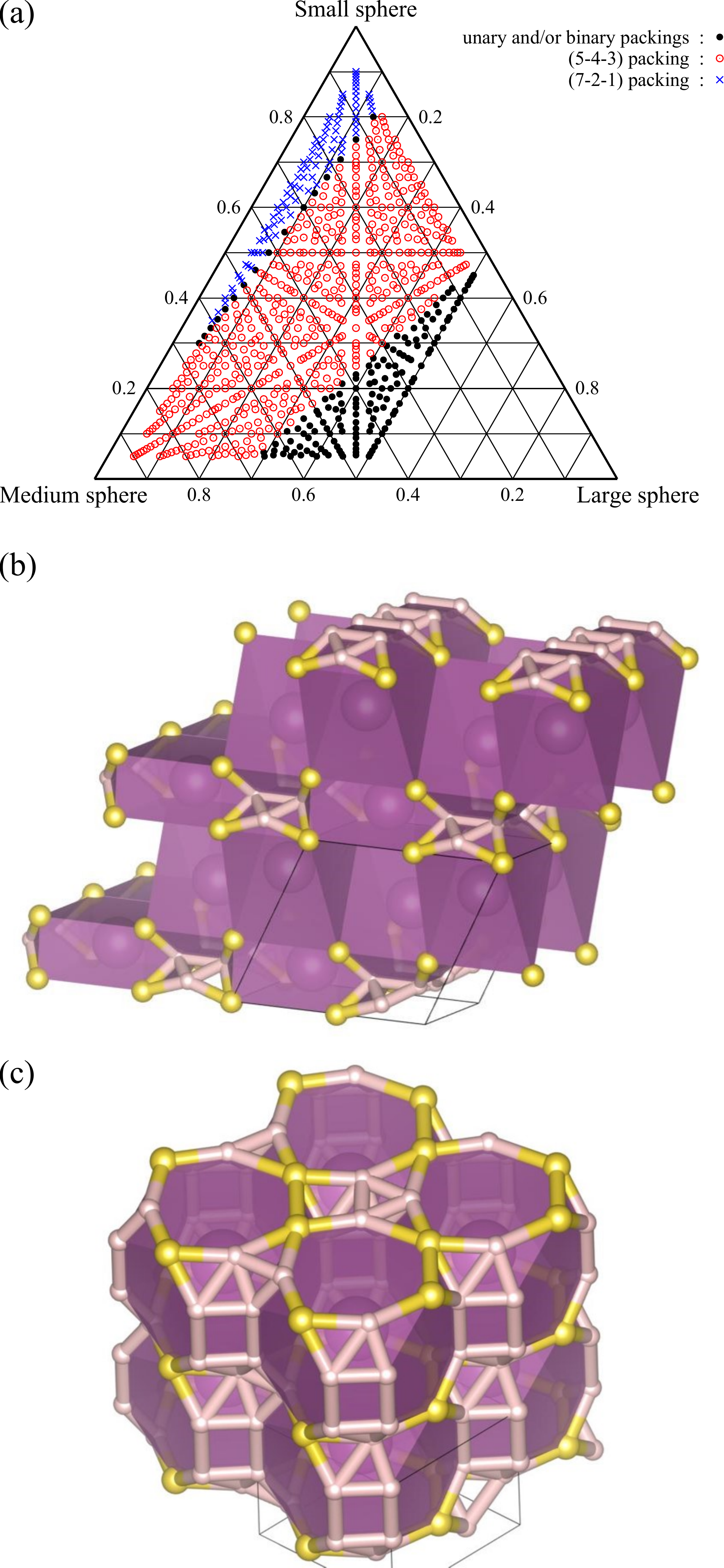}
\caption{The phase diagram at the radius ratio of $0.34:0.44:1.00$ (a) shows that the (5-4-3) structure (b) and the (7-2-1) structure (c) are the DTSPs. The packing fractions of these two DTSPs are $0.768812$ and $0.769884$, respectively. The (5-4-3) structure, which has already been discussed in the previous study~\cite{PhysRevE.104.024101} as SDTSP, has the $Imm2$ symmetry if the slight distortion is corrected. Besides, if the structural distortion of the (7-2-1) structure is corrected, it has the $Amm2$ symmetry.}
\label{fig:034-044-100}
\end{figure}
\begin{figure}
\centering
\includegraphics[width=0.65\columnwidth]{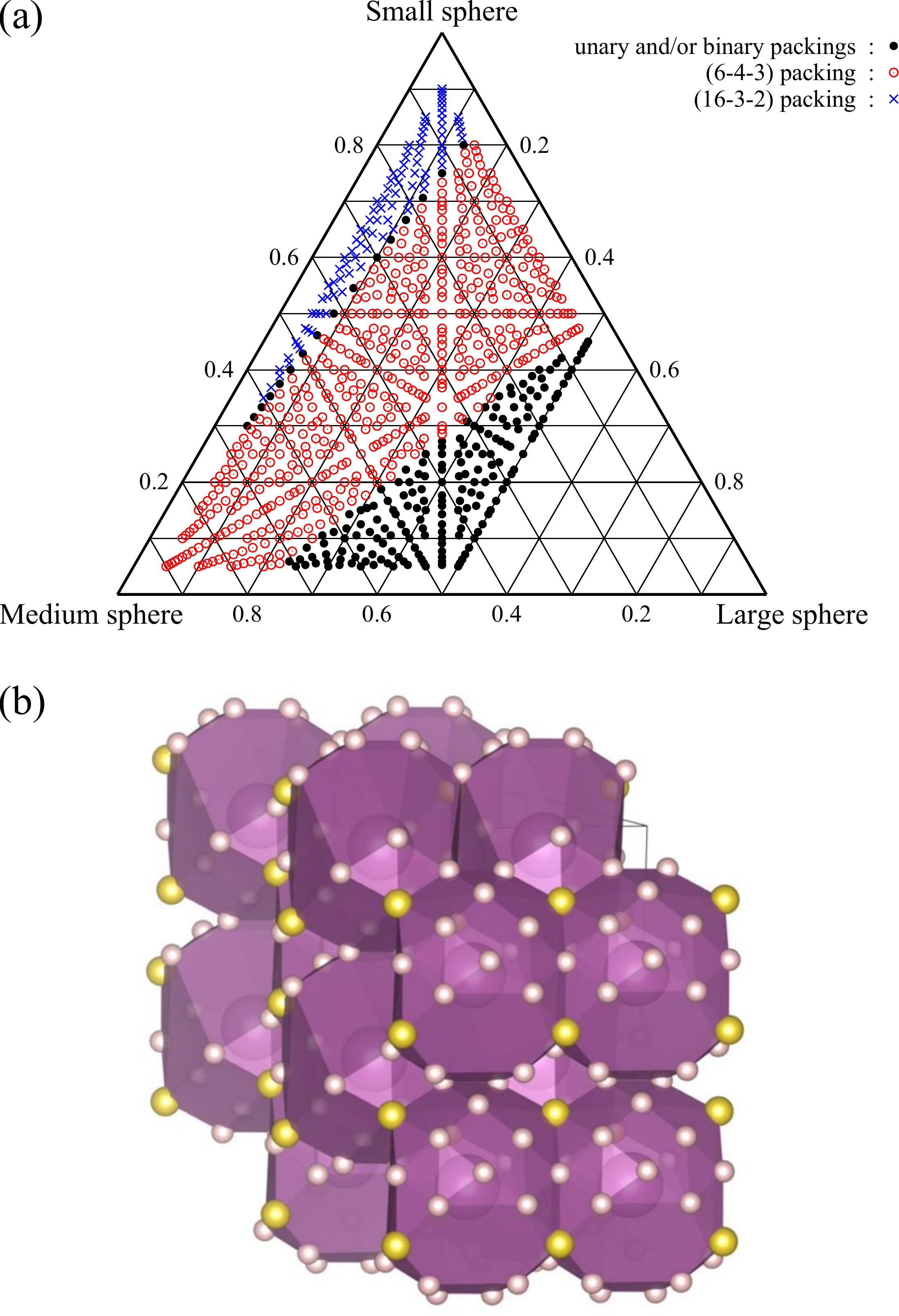}
\caption{The phase diagram at the radius ratio of $0.34:0.45:1.00$ (a) The (16-3-2) structure (b) is the DTSP, and the packing fraction is $0.769635$. The (6-4-3) structure plotted on the phase diagram (a) has the packing fraction of $0.771870$, and the structure is shown in Fig.~\ref{fig:033-044045-100}(c).}
\label{fig:034-045-100}
\end{figure}
\begin{figure}
\centering
\includegraphics[width=0.65\columnwidth]{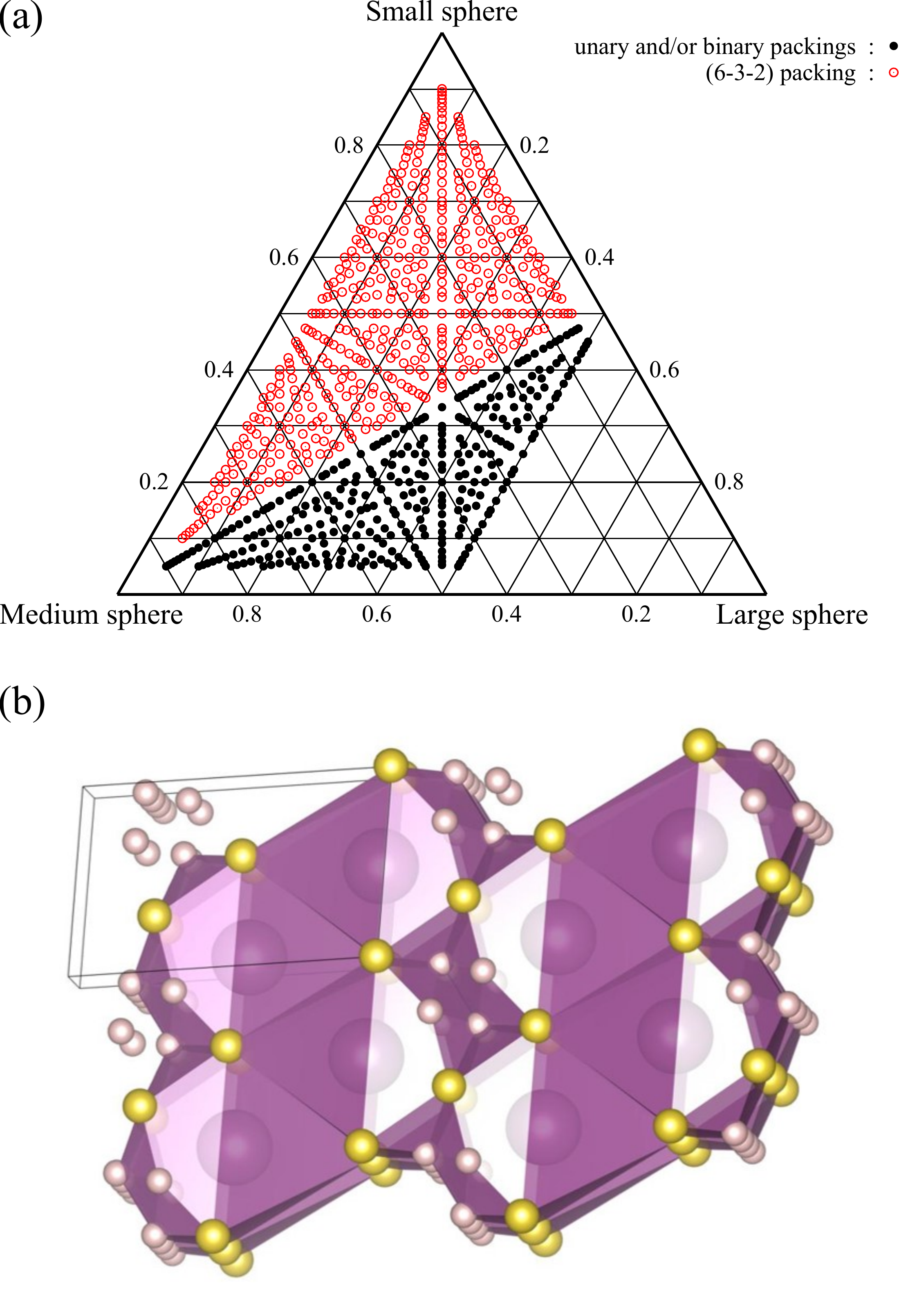}
\caption{The phase diagram at the radius ratio of $0.35:0.46:1.00$ (a) shows that the (6-3-2) structure (b) is the DTSP, and the packing fraction is $0.766820$.}
\label{fig:035-046-100}
\end{figure}
\begin{figure}
\centering
\includegraphics[width=0.65\columnwidth]{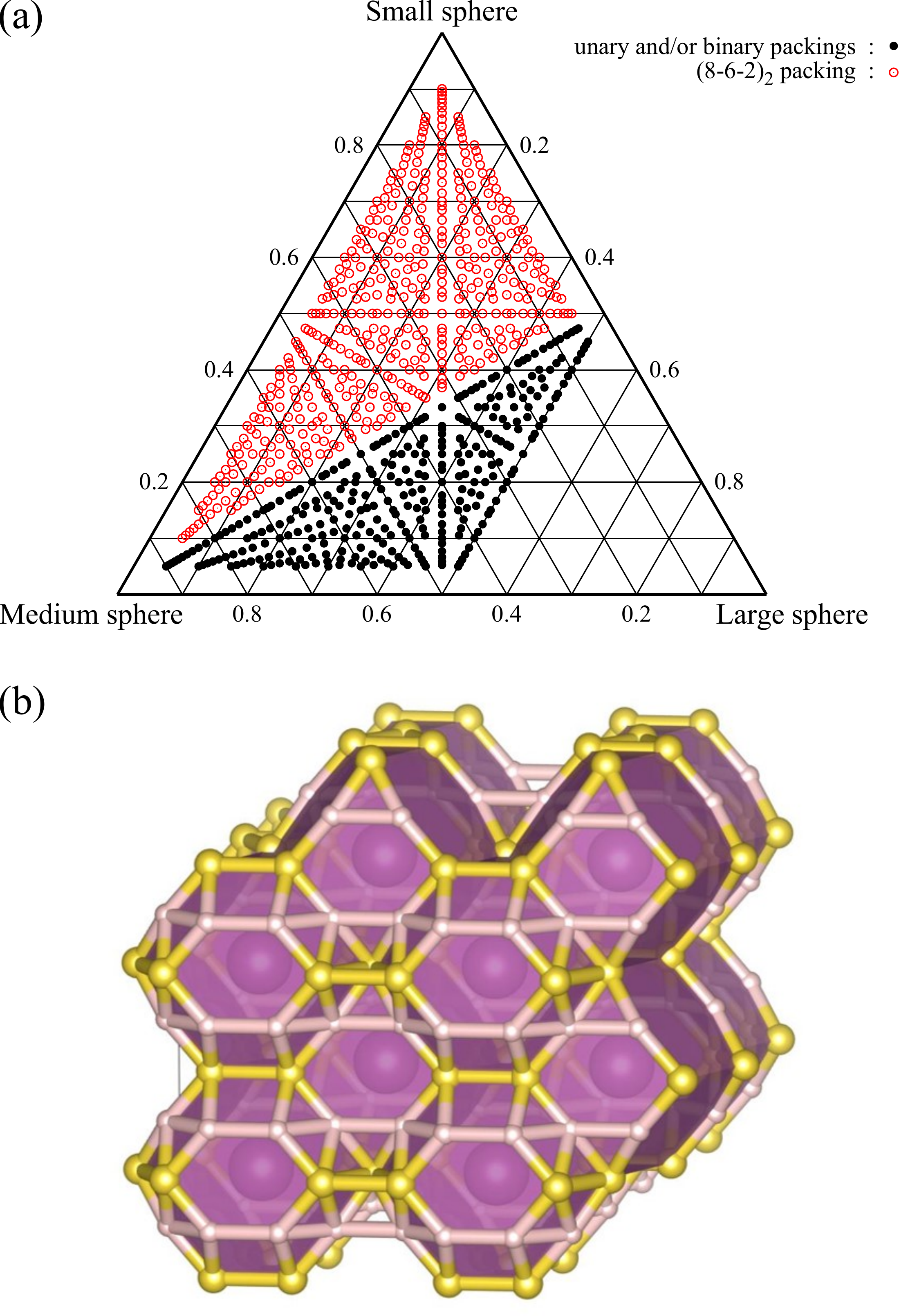}
\caption{The phase diagram at the radius ratio of $0.35:0.48:1.00$ (a) show that the (8-6-2)$_2$ structure (b) is the DTSP, and the packing fraction is $0.764043$. If the structural distortion is corrected, this structure has the $Pmmn$ symmetry.}
\label{fig:035-048-100}
\end{figure}
\begin{figure}
\centering
\includegraphics[width=0.65\columnwidth]{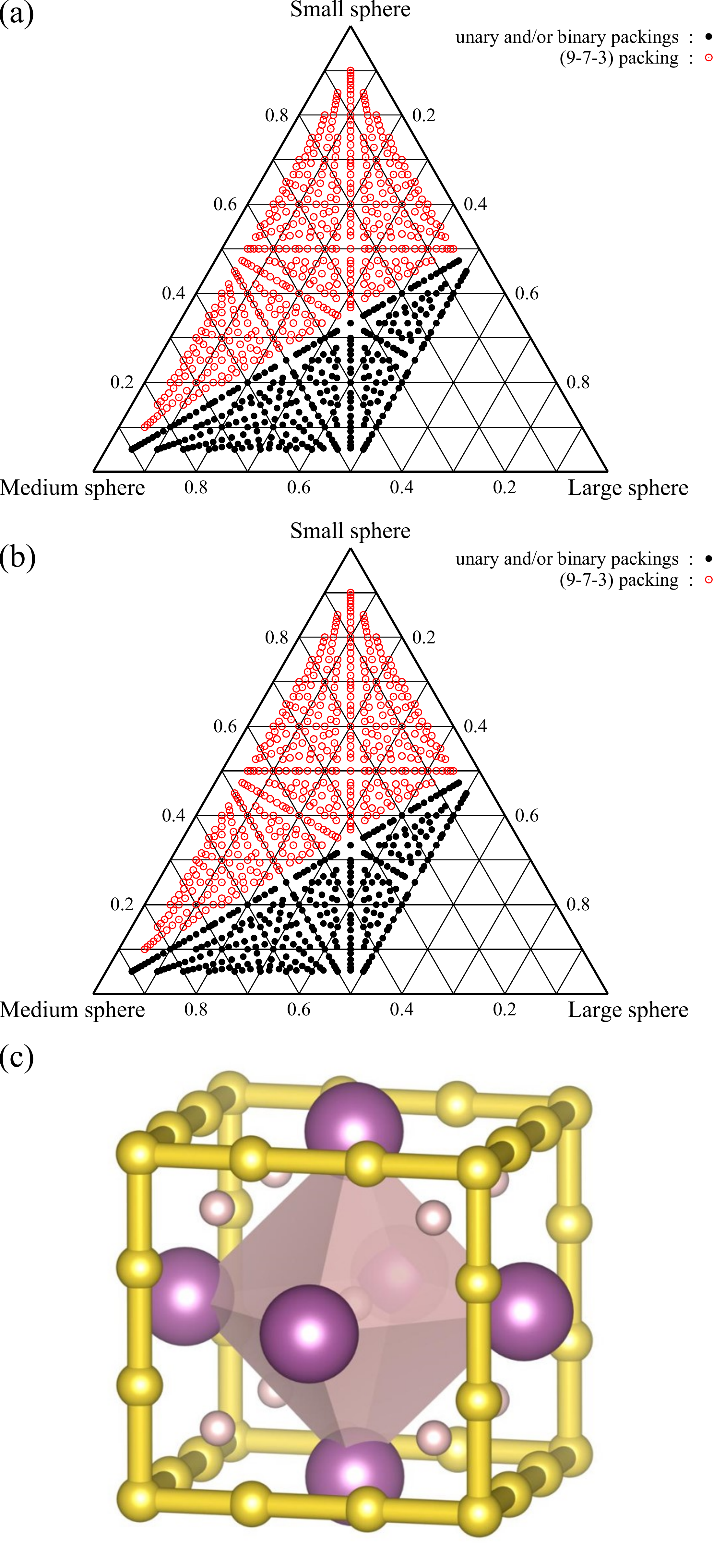}
\caption{The phase diagrams at the radius ratios of $0.36:0.47:1.00$ (a) and $0.37:0.47:1.00$ (b) shows that the (9-7-3) structure (c) is the DTSP, and the packing fractions at the two radius ratios are $0.767631$ and $0.774290$, respectively. If the structural distortion is corrected, this structure has the $Pm \bar{3}m$ symmetry.}
\label{fig:036037-047-100}
\end{figure}
\begin{figure}
\centering
\includegraphics[width=0.65\columnwidth]{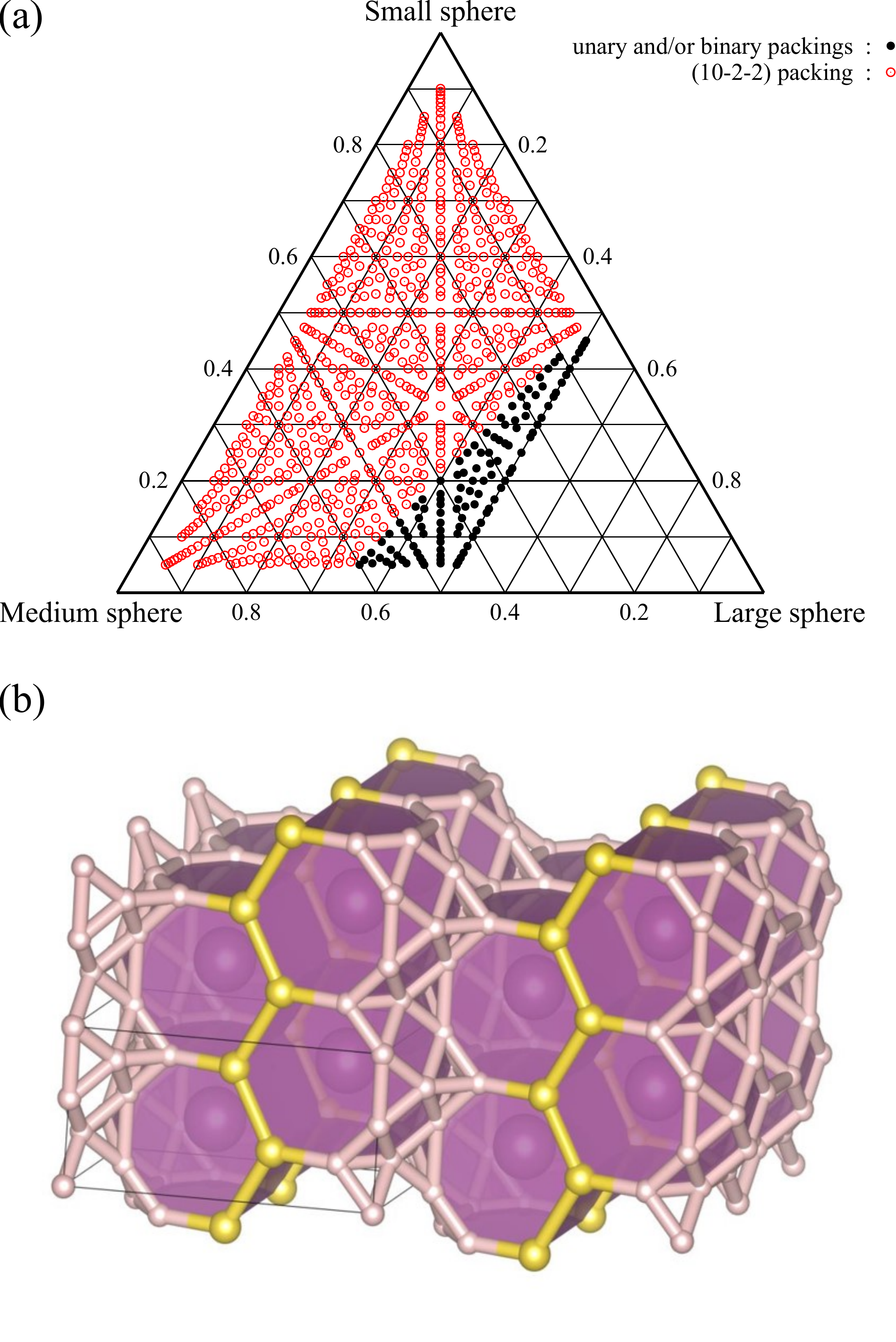}
\caption{The phase diagram at the radius ratio of $0.38:0.52:1.00$ (a) shows that the (10-2-2) structure (b) is the DTSP, and the packing fraction is $0.772794$. If the structural distortion is corrected, this structure has the $Cmcm$ symmetry.}
\label{fig:038-052-100}
\end{figure}
\begin{figure}
\centering
\includegraphics[width=0.65\columnwidth]{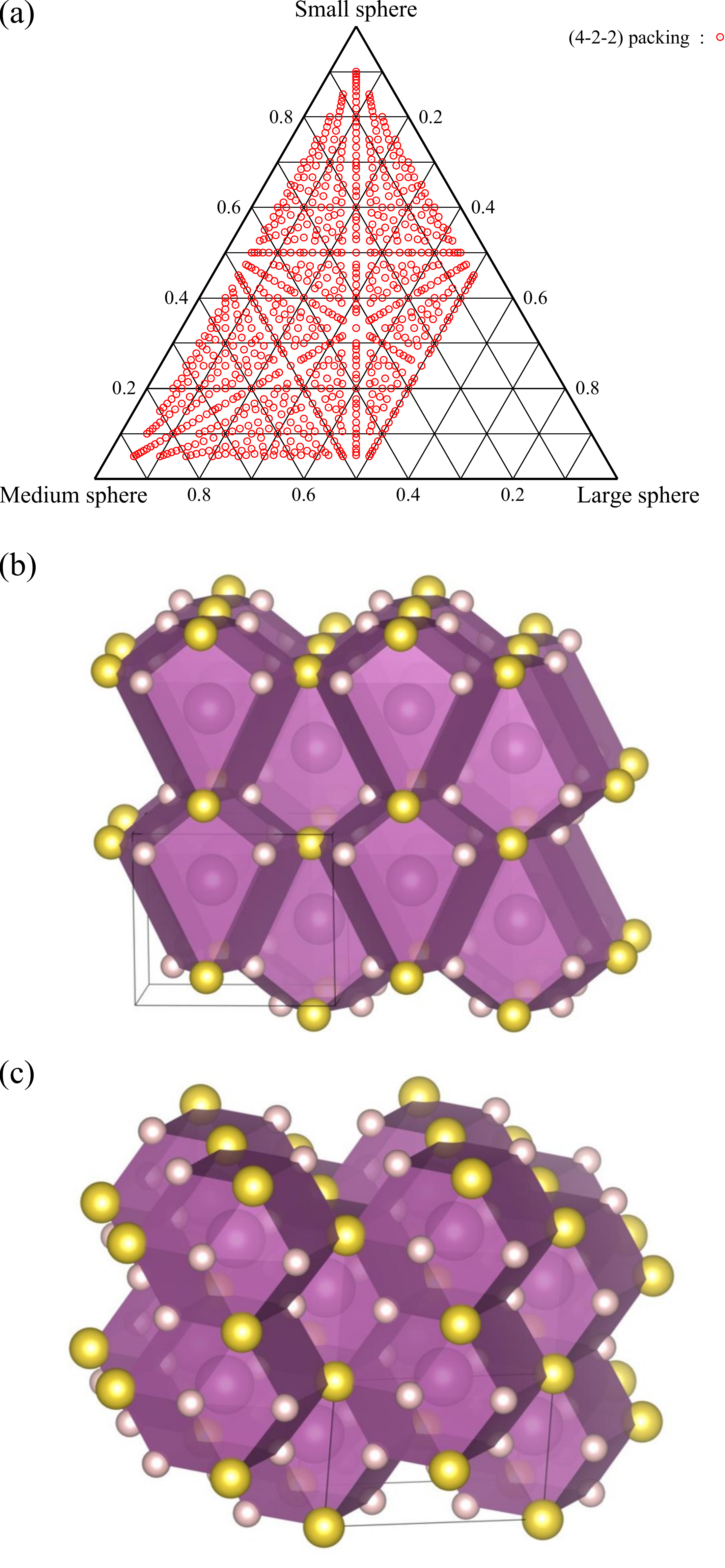}
\caption{The common phase diagram (a) at the seven radius ratios of $0.48:0.64:1.00$, $0.48:0.66:1.00$, $0.48:0.68:1.00$, $0.48:0.70:1.00$, $0.48:0.72:1.00$, $0.50:0.66:1.00$, and $0.50:0.68:1.00$. shows that the (4-2-2)$\mathrm{_5}$ structure (b) is the DTSP, and the packing fractions at the seven radius ratios are $0.747827$, $0.758687$, $0.761453$, $0.758173$, $0.756223$, $0.758050$, and $0.769880$, repsectively. This structure, which has already been discussed in the previous study~\cite{PhysRevE.104.024101} as SDTSP, has the $Pmmn$ symmetry if the structural distortion is corrected. At the three radius ratios of $0.48:0.68:1.00$, $0.48:0.70:1.00$, and $0.48:0.72:1.00$, the (4-2-2)$\mathrm{_6}$ structure(c) has the same packing fractions as those of the (4-2-2)$\mathrm{_5}$ structure. This structure, which has also already been discussed in the previous study~\cite{PhysRevE.104.024101} as SDTSP, has the $R \bar{3}m$ symmetry if the structural distortion is corrected. Note that the (4-2-2)$\mathrm{_6}$ structure is equal to the (2-1-1) structure, which was found in the previous study~\cite{PhysRevE.104.024101}, if the structural distortion in the (4-2-2)$\mathrm{_6}$ structure is corrected enough to reduce the unit cell.}
\label{fig:largeDTSPs}
\end{figure}

\begin{table}
\caption{The number of generated structures for exploring DTSPs at each composition.}
\label{table:number-of-genetated-structures}
\begin{ruledtabular}
\begin{tabular}{cc}
Range of $N$ & Number of generated structures \\
$3 \le N \le 5$ & 200,000 \\
$6 \le N \le 7$ & 700,000 \\
$8 \le N \le 10$ & 2,000,000 \\
$11 \le N \le 15$ & 10,000,000 \\
$16 \le N \le 20$ & 20,000,000 \\
$21 \le N \le 25$ & 40,000,000 \\
$26 \le N \le 30$ & 60,000,000 \\
\end{tabular}
\end{ruledtabular}
\end{table}

\begin{table}
\caption{The (2-2-2)$_1$ structure~\cite{PhysRevE.104.024101} shown in Fig.~\ref{fig:knownDTSPs}(a) appears on the phase diagram at the seven radius ratios.}
\label{table:radius_ratios_of_2-2-2_1}
\begin{ruledtabular}
\begin{tabular}{ccc}
Radius ratio & Radius ratio & Radius ratio \\
\hline
$0.29:0.45:1.00$ & $0.29:0.46:1.00$ & $0.30:0.45:1.00$ \\
$0.30:0.46:1.00$ & $0.31:0.45:1.00$ & $0.31:0.47:1.00$ \\
$0.32:0.47:1.00$ & & \\
\end{tabular}
\end{ruledtabular}
\end{table}
\begin{table}
\caption{The (2-2-2)$_2$ structure~\cite{PhysRevE.104.024101} shown in Fig.~\ref{fig:knownDTSPs}(b) appears on the phase diagram at the six radius ratios.}
\label{table:radius_ratios_of_2-2-2_2}
\begin{ruledtabular}
\begin{tabular}{ccc}
Radius ratio & Radius ratio & Radius ratio \\
\hline
$0.29:0.49:1.00$ & $0.29:0.50:1.00$ & $0.29:0.51:1.00$ \\
$0.30:0.49:1.00$ & $0.30:0.50:1.00$ & $0.31:0.49:1.00$ \\
\end{tabular}
\end{ruledtabular}
\end{table}
\begin{table}
\caption{The (4-3-1) structure~\cite{PhysRevE.104.024101} has the $P6/mmm$ symmetry if the structural distortion is corrected. Medium spheres constitute the kagome lattice, as shown in Fig.~\ref{fig:knownDTSPs}(c). This structure appears on the phase diagrams at the 22 radius ratios.}
\label{table:radius_ratios_of_4-3-1}
\begin{ruledtabular}
\begin{tabular}{ccc}
Radius ratio & Radius ratio & Radius ratio \\
\hline
$0.29:0.42:1.00$ & $0.29:0.43:1.00$ & $0.29:0.44:1.00$ \\
$0.29:0.45:1.00$ & $0.29:0.46:1.00$ & $0.29:0.47:1.00$ \\
$0.29:0.48:1.00$ & $0.29:0.49:1.00$ & $0.29:0.50:1.00$ \\
$0.30:0.43:1.00$ & $0.30:0.44:1.00$ & $0.30:0.45:1.00$ \\
$0.30:0.46:1.00$ & $0.30:0.47:1.00$ & $0.30:0.48:1.00$ \\
$0.30:0.49:1.00$ & $0.30:0.50:1.00$ & $0.31:0.47:1.00$ \\
$0.31:0.48:1.00$ & $0.31:0.49:1.00$ & $0.31:0.50:1.00$ \\
$0.31:0.51:1.00$ & & \\
\end{tabular}
\end{ruledtabular}
\end{table}
\begin{table}
\caption{The (4-4-2) structure~\cite{PhysRevE.104.024101}, which can be directly derived from the $\mathrm{HgBr}_2$ structure as shown in Fig.~\ref{fig:knownDTSPs}(d), has the $Cmcm$ symmetry if the structural distortion is corrected. This structure appears on the phase diagram at the 16 radius ratios.}
\label{table:radius_ratios_of_4-4-2}
\begin{ruledtabular}
\begin{tabular}{ccc}
Radius ratio & Radius ratio & Radius ratio \\
\hline
$0.29:0.42:1.00$ & $0.29:0.43:1.00$ & $0.29:0.44:1.00$ \\
$0.29:0.45:1.00$ & $0.29:0.46:1.00$ & $0.29:0.47:1.00$ \\
$0.30:0.42:1.00$ & $0.30:0.43:1.00$ & $0.30:0.44:1.00$ \\
$0.30:0.45:1.00$ & $0.30:0.46:1.00$ & $0.30:0.47:1.00$ \\
$0.31:0.43:1.00$ & $0.31:0.44:1.00$ & $0.31:0.45:1.00$ \\
$0.31:0.47:1.00$ & & \\
\end{tabular}
\end{ruledtabular}
\end{table}
\begin{table}
\caption{The unit cell of the (10-6-3) structure~\cite{PhysRevE.104.024101} are comprised of medium spheres as shown in \ref{fig:knownDTSPs}(f). The (10-6-3) structure has the $Pm\bar{3}m$ symmetry if the structural distortion is corrected. This structure appears on the phase diagrams at the seven radius ratios.}
\label{table:radius_ratios_of_10-6-3}
\begin{ruledtabular}
\begin{tabular}{ccc}
Radius ratio & Radius ratio & Radius ratio \\
\hline
$0.33:0.52:1.00$ & $0.34:0.50:1.00$ & $0.34:0.51:1.00$ \\
$0.35:0.49:1.00$ & $0.35:0.50:1.00$ & $0.35:0.51:1.00$ \\
$0.36:0.49:1.00$ & & \\
\end{tabular}
\end{ruledtabular}
\end{table}
\begin{table}
\caption{The (13-2-1) structure, which is shown in in Fig.~\ref{fig:knownDTSPs}(g), has the $Fm \bar{3}m$ symmetry if the structural distortion is corrected. The structure appears on the phase diagram at the eight radius ratios listed in Table \ref{table:radius_ratios_of_13-2-1}. Note that for $0.40 \le \alpha_1 \le 0.64$, we choose only 13 values of $0.40, 0.42, \cdots 0.64$, and let $\alpha_2$ be chosen only from 18 values from $0.50, 0.52, \cdots 0.74$, as discussed in Sec.~\ref{sec:compositions}.}
\label{table:radius_ratios_of_13-2-1}
\begin{ruledtabular}
\begin{tabular}{ccc}
Radius ratio & Radius ratio & Radius ratio \\
\hline
$0.42:0.62:1.00$ & $0.42:0.64:1.00$ & $0.44:0.62:1.00$ \\
$0.44:0.64:1.00$ & $0.44:0.66:1.00$ & $0.46:0.60:1.00$ \\
$0.46:0.62:1.00$ & $0.46:0.64:1.00$ & \\
\end{tabular}
\end{ruledtabular}
\end{table}
\begin{table}
\caption{The (13-3-1) structure, which is shown in in Fig.~\ref{fig:knownDTSPs}(h), has the $Pm \bar{3}m$ symmetry if the structural distortion is corrected. This structure appears on the phase diagram at the nine radius ratios.}
\label{table:radius_ratios_of_13-3-1}
\begin{ruledtabular}
\begin{tabular}{ccc}
Radius ratio & Radius ratio & Radius ratio \\
\hline
$0.29:0.41:1.00$ & $0.29:0.42:1.00$ & $0.29:0.43:1.00$ \\
$0.30:0.40:1.00$ & $0.30:0.41:1.00$ & $0.30:0.42:1.00$ \\
$0.30:0.43:1.00$ & $0.31:0.41:1.00$ & $0.31:0.42:1.00$ \\
\end{tabular}
\end{ruledtabular}
\end{table}

In this section, we summarize the method to explore the DTSPs and describe the exploration areas including radius ratios and compositions.

\subsection{Exploration method}

Our previous study~\cite{PhysRevE.103.023307, PhysRevE.104.024101} showed that the algorithm, which is based on the piling-up and iterative-balance methods, is effective enough to search the densest sphere packings (DSPs). The piling-up method is an approach to randomly generate initial structures for exploring DSPs. The random approach, which directly generates multilayered structures with modest overlaps, enables us to find a wide variety of packings due to the unbiased distribution of initial structures in the configuration space. The iterative balance method optimizes initial structures to dense packing structures by minimizing the volume of unit cells and the overlaps between spheres simultaneously by the steepest descent method under pressure with the hard-sphere potential. The method can find the most optimal structural distortion for high density since the pressure makes as many distances between spheres as possible converge to zero. Although a large number of optimization steps is necessary to find the local highest packing fractions accurately, the computational cost is low because of the efficiency of each step. Before the fine iterative balance optimization, we apply the pseudoannealing to optimize initial structures to packing structures with overlap. Besides, to reduce the computational cost, coarse iterative balance optimization is applied to reject sparse structures based on the rough estimation of the packing fraction. In this study, we apply the algorithm to the detailed exploration of the DTSPs, where all parameters of the exhaustive exploration and reoptimization for DTSPs are set to the default values given in Ref.~\cite{PhysRevE.104.024101}.

\subsection{Exploration conditions}
\label{sec:exploration_conditions}

We have exhaustively explored DTSPs at 451 kinds of radius ratios and 436 kinds of compositions. In this subsection, we detail the conditions.

\subsubsection{Radius Ratios}

We write $\alpha_1$ and $\alpha_2$ as the radii of small and medium spheres, respectively, where we fix the radius of large spheres to be $1.0$. 
For $\alpha_1$ and $\alpha_2$, we impose a restriction as 
\begin{equation}
\alpha_1 + 0.1 \le \alpha_2. \label{eq:radii-condition-1}
\end{equation}
The radius ratios at which we explore the DTSPs are given below:
\begin{itemize}
\item For $0.29 \le \alpha_1 \le 0.38$, we choose ten values of $0.29, 0.30, \cdots 0.38$ for $\alpha_1$. Letting $\alpha_2$ be chosen from 36 values from $0.39, 0.40, \cdots 0.74$ under the constraint to $\alpha_1$ and $\alpha_2$ given in Eq.~(\ref{eq:radii-condition-1}), the total number of radius ratios is 315.
\item For $0.40 \le \alpha_1 \le 0.64$, we choose 13 values of $0.40, 0.42, \cdots 0.64$ for $\alpha_1$. Letting $\alpha_2$ be chosen from 18 values from $0.50, 0.52, \cdots 0.74$ under the constraint to $\alpha_1$ and $\alpha_2$ given in Eq.~(\ref{eq:radii-condition-1}), the total number of radius ratios is 91.
\item For $0.66 \le \alpha_1 \le 0.82$, we choose nine values of $0.66, 0.68, \cdots 0.82$ for $\alpha_1$. Letting $\alpha_2$ be chosen from nine values from $0.76, 0.78, \cdots 0.92$ under the constraint to $\alpha_1$ and $\alpha_2$ given in Eq.~(\ref{eq:radii-condition-1}), the total number of radius ratios is 45.
\end{itemize}

\subsubsection{Compositions}
\label{sec:compositions}

We write $n_1$, $n_2$, and $n_3$ as the numbers of small, medium, and large spheres per unit cell, respectively; the total number of spheres per unit cell is denoted by $N$. We explore the DTSPs for all compositions which satisfy the three constraints shown below:
\begin{equation}
3 \le N \le 25, \label{eq:total-number-of-n1-n2-n3}
\end{equation}
\begin{equation}
n_3 \le n_2 \le n_1, \label{eq:constraint-of-n1-n2-n3-1}
\end{equation}
\begin{equation}
n_1 \le 5 \left(n_2 + n_3 \right). \label{eq:constraint-of-n1-n2-n3-2}
\end{equation}
The number of compositions which satisfy Eqs.~(\ref{eq:total-number-of-n1-n2-n3}), (\ref{eq:constraint-of-n1-n2-n3-1}), and (\ref{eq:constraint-of-n1-n2-n3-2}) is 436. Note that the third constraint eliminates compositions which are comprised by a relatively large number of small spheres. The exceptional cases are given below:
\begin{itemize}
\item For $0.40 \le \alpha_1 \le 0.64$, $N$ is set to be from 3 to 30. In this case, the number of compositions becomes 743. 
\item For $\alpha_1 = 0.29$, $N$ is set to be from 3 to 20. In this case, the number of compositions becomes 227.
\end{itemize}
The number of generated structures at each composition in creating initial structures is given in Table~\ref{table:number-of-genetated-structures}.

Note that the discovered DTSPs are named as ($n_1$-$n_2$-$n_3$) structure. When the two DTSPs have the same composition, we distinguish them by index.

\subsubsection{Phase Separation}

We denote the composition ratios of small, medium, and large spheres as $x_1$, $x_2$, and $x_3$, respectively, which are defined as
\begin{equation}
x_i \equiv \frac{n_i}{N}. \label{eq:definition-of-composition-ratio-xi}
\end{equation}
As discussed in previous studies~\cite{PhysRevE.85.021130, PhysRevE.103.023307}, for a given composition ratio $\left(x_1, x_2, x_3\right)$ of three kinds of spheres, the densest packing fraction can be achieved by the phase separation which consists of less than or equal to three structures when every candidate structure is periodic. Therefore, we must construct the phase diagrams, which show the densest phase separation at each radius and composition ratio, so that we identify the densest multiary sphere packings. Note that we must know the accurate packing fractions of DBSPs at each radius ratio since the consideration of the phase separation including the DBSPs is necessary to get the maximum packing fraction at each radius and composition ratio. For ease of visibility, only the 779 composition ratios that fullfill the conditions:
\begin{equation}
n_3 \le n_1 + n_2, 3 \le N \le 20, \label{eq:constraint-of-n1-n2-n3-4}
\end{equation}
are plotted on the phase diagrams shown in this study.

\section{Result and Discussion}
\label{sec:results_and_discussion}

\begin{figure*}
\centering
\includegraphics[width=1.8\columnwidth]{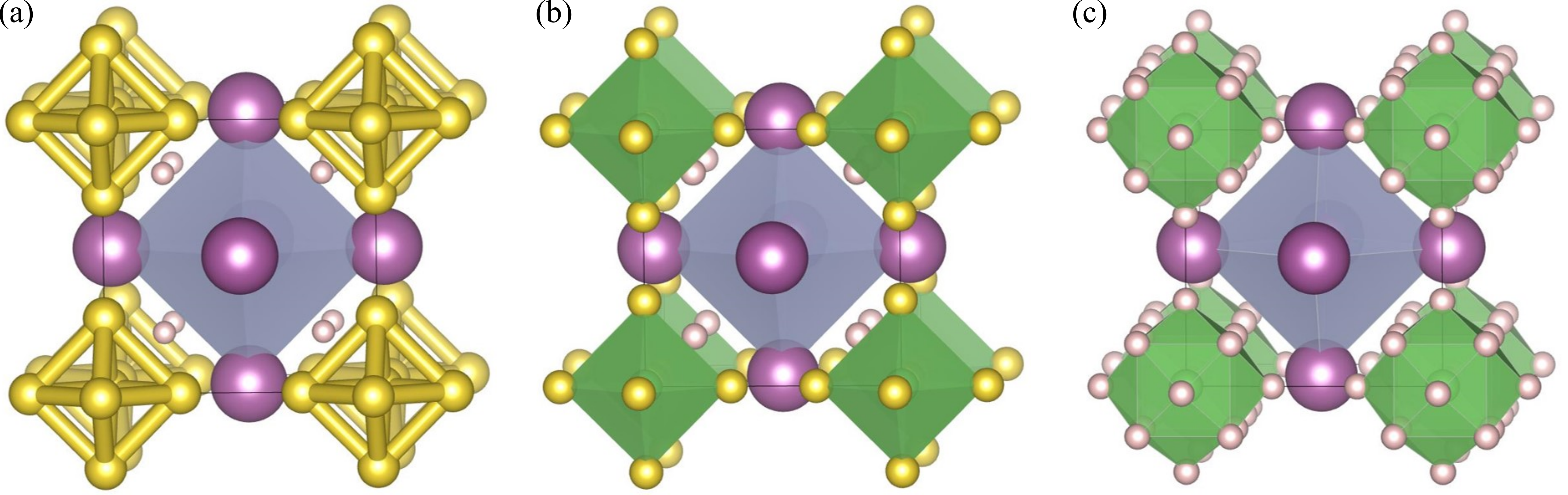}
\caption{(a) The (8-1-6-3) structure, in which a fourth sphere replaces one small sphere at the center of the unit cell of the (9-6-3) structure shown in Fig.~\ref{fig:knownDTSPs}(e). (b) The (8-1-6-1-3) structure, in which a fourth sphere replaces one small sphere at the center of the unit cell of the (9-7-3) structure shown in Fig.~\ref{fig:036037-047-100}(c), and a fifth sphere replaces one medium sphere at the vertex of the unit cell. (c) The (14-1-1-3) structure, in which all the medium spheres in the (8-1-6-1-3) structure is replaced by small spheres.}
\label{fig:prototypes}
\end{figure*}
\begin{figure}
\centering
\includegraphics[width=0.5\columnwidth]{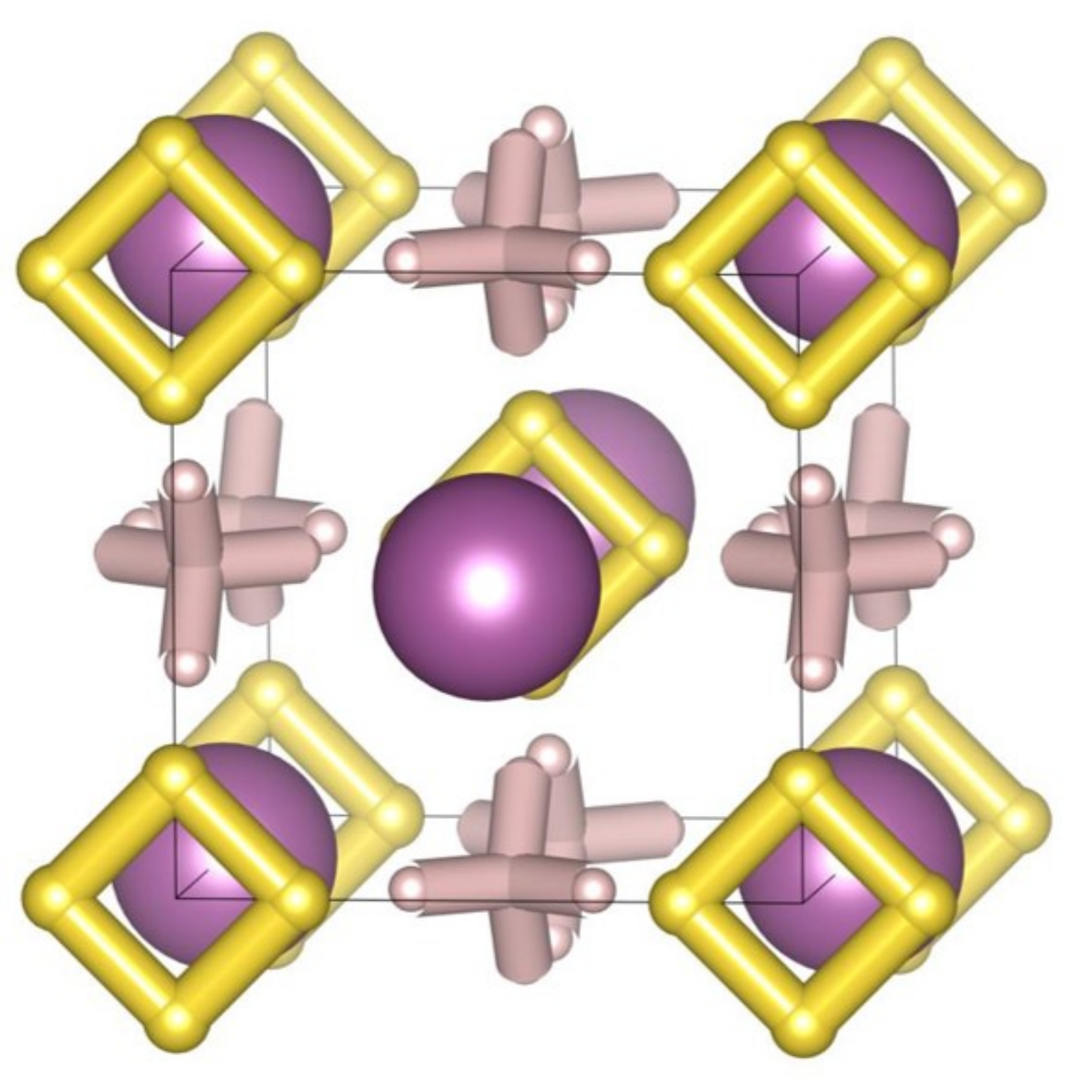}
\caption{The (10-4-1) structure that has already been discovered in the previous study~\cite{PhysRevE.104.024101}.  A tetrahedral site comprised of large spheres is occupied by a tetrahedron consisting of five small spheres, where one small sphere is placed at the center of the tetrahedron.}
\label{fig:10-4-1}
\end{figure}

In this section, we discuss the discovered DTSPs, constructed phase diagrams, the geometric features of the DTSPs, and the correspondence between DTSPs and crystals.

\subsection{DTSPs on phase diagrams}

In our exhaustive search for the DTSPs, we find the 22 kinds of unknown DTSPs including the six structures identified as SDTSPs in Ref.~\cite{PhysRevE.104.024101}, and the (8-6-2)$_{\mathrm{L}}$ structure that is the long-period structure of the (4-3-1) structure shown in Fig.~\ref{fig:knownDTSPs}(c). The discovered DTSPs are shown in Figs.~\ref{fig:029-039-100} to \ref{fig:largeDTSPs} with the corresponding phase diagrams and the captions describing the basic structural information such as packing fraction and spacegroup. In all the figures showing structures, the lines between spheres do not necessarily correspond to the contact between spheres. The spacegroup is determined by the code Spglib~\cite{togo2018textttspglib} with correcting the structural distortions in each DTSP. Three-dimensional data of the discovered DTSPs are available online~\cite{samlai}. Note that it is not apparent whether there is at least one DTSP in the densest phase separation at any radius and composition ratio, since the densest packing fraction may consist of only the FCC structures and/or DBSPs.

The discovered DTSPs tend to be well-ordered, for example, the (9-7-3) structure shown in Fig.~\ref{fig:036037-047-100}(c) has the $Pm \bar{3}m$ symmetry if the distortion is corrected. Four medium spheres are placed on the edge of the cubic unit cell, one large sphere is placed at the center of each surface of the unit cell, and nine small spheres are placed into the cubic cell, and one of them is placed at the center. This structural property is similar to that of the (9-6-3) and (10-6-3) structures~\cite{PhysRevE.104.024101} shown in Figs.~\ref{fig:knownDTSPs}(e) and \ref{fig:knownDTSPs}(f). The difference between the (9-6-3) and (9-7-3) structures is whether or not one medium sphere is placed at the vertex of the unit cell, while in (10-6-3) structure, one small sphere is placed at the vertex of the unit cell. These three packings are related to the perovskite structure, and in this sense, they are similar to the (13-3-1) structure shown in Fig.~\ref{fig:knownDTSPs}(h). The small difference of the radius ratios makes a change that which DTSPs are the densest.

As the (9-7-3) structure, some of the discovered DTSPs are relevant to the other DBSPs and DTSPs, for example, the (6-6-4) structure shown in Fig.~\ref{fig:029-048-100} is composed of the local structures in the (6-1) and (16-4) structures that are the two of the DBSPs~\cite{PhysRevLett.107.125501, PhysRevE.85.021130, PhysRevE.103.023307}. In fact, small spheres constitute a chain structure that can be seen in the (6-1) structure, and medium spheres constituting an octahedron are placed in a distorted rectangular consisting of large spheres as in the (16-4) structure. These local structures can be seen in crystals; the (16-4) structure corresponds to the crystal of the $\mathrm{UB}_4$, and the (6-1) structure corresponds to the crystal of the $\mathrm{YH}_6$ under high pressure~\cite{https://doi.org/10.1002/adma.202006832}, as discussed in Ref.~\cite{PhysRevE.103.023307}. Therefore, we can expect that the (6-6-4) structure might be realized by materials under high pressure. The chain structure of small spheres can also be seen in the (10-3-2) and (6-2-2) structure shown in Figs.~\ref{fig:029-050-100} and \ref{fig:029-051-100}, respectively. Besides, the kagome lattice of medium spheres can be seen in the (10-3-2) structure as well as the (4-3-1) structure shown in Fig.~\ref{fig:knownDTSPs}(c), and honeycomb lattice can be seen in not only the (6-2-2) structure but also the $\mathrm{AuTe}_2$ structure that is one of the DBSP~\cite{PhysRevE.79.046714, PhysRevLett.107.125501, PhysRevE.85.021130}. Finally, the difference between the (5-4-3) and (6-4-3) structures shown in Figs.~\ref{fig:033-044045-100} and \ref{fig:034-044-100} comes from the absence or existence of one small sphere between the two medium spheres in the cluster. The resemblance among DTSPs indicates that there are dense local structures as discussed in the binary case~\cite{PhysRevE.103.023307}. Whether or not the local structures comprise the DTSPs is equivalent to whether or not it is possible to combine some local structures with periodicity so that there are no unnecessary voids or distortions.

This study also shows that the well-ordered DTSPs which have already been discovered in the previous study~\cite{PhysRevE.104.024101} appears on the ternary phase diagrams at many radius ratios. Tables \ref{table:radius_ratios_of_2-2-2_1} to \ref{table:radius_ratios_of_13-3-1} show the radius ratios at which the seven of the nine DTSPs shown in Fig.~\ref{fig:knownDTSPs} appear on the phase diagrams. In addition, the (9-6-3) and (16-2-2) structures shown in Figs.~\ref{fig:knownDTSPs}(e) and \ref{fig:knownDTSPs}(i), respectively, appear on the phase diagram at the radius ratio of $0.30:0.55:1.00$ and $0.35:0.45:1.00$, respectively. Finally, the (2-1-1) structure~\cite{PhysRevE.104.024101}, whose unit cell is half of that of the corrected (4-2-2)$_6$ structure, appears on the phase diagrams at the two radius ratios of $0.48:0.74:1.00$ and $0.50:0.70:1.00$.

\subsection{Full pictures of the ternary phase diagrams}

In this study, we have exhaustively explored the DTSPs at 451 kinds of radius ratios, however, at a considerable number of radius ratios, the highest packing fractions are achieved by the phase separations consisting of only the FCC and/or the DBSPs for all compositions; the DTSPs appear at only 76 kinds of radius ratios. The tendency is getting evident as the small and medium spheres are getting larger, in fact, we find no DTSPs for the case of $0.66 \le \alpha_1 \le 0.82$. Based on the hypothesis that the DTSPs are composed of the combination of some local structures with dense so that there are no unnecessary voids or distortions, the tendency may indicate an assumption that the dense local structures might be getting complex as the small and medium spheres are getting larger. Therefore, we cannot exclude the possibility that the unit cell of the undiscovered DTSPs might be much larger than in this study, especially in the case that the radii of small and medium spheres are relatively large. Of course, the local structures in the DBSPs may be denser than those consisting of three kinds of spheres; the DBSPs may have higher packing fractions than any other ternary packings. Further study is necessary to determine which hypothesis is true.

On the other hand, in the case that $\alpha_1 = 0.29$, we find nine DTSPs. The discovery indicates that the phase diagrams may be getting complex as the radii of small and medium spheres are getting smaller, in fact, the previous study~\cite{PhysRevE.104.024101} showed that there are seven putative DTSPs at the radius ratios of $0.20:0.45:1.00$. Based on the hypothesis that as the radii of small and medium spheres are getting small, the small and medium spheres are getting easier to be placed in small voids consisting of large spheres and besides the clusters of them is also getting easier to be placed with large spheres, the tendency may indicate an assumption that there are many competitive structures with respect to the packing fraction in this area. The competition may cause the complexity of phase diagrams; there might be unknown DTSPs that appear on the phase diagrams in a very narrow range of radius ratio as the (7-3) structure~\cite{PhysRevLett.107.125501, PhysRevE.85.021130, PhysRevE.103.023307}. For the radius ratios of $\alpha < 0.30$, more detailed research with respect to the radius ratios is necessary to get full pictures of the ternary phase diagrams in this area.

\subsection{Geometric features of discovered DTSPs}
\label{sec:geometric_features_of_discovered_dtsps}

We analyze the structural properties, and accordingly, we classify the DTSPs by how the structural framework is constituted.

The structural framework of the (9-7-3) structure is composed of medium spheres as the cubic unit cell of the (9-7-3) structure is constituted by only seven medium spheres, as shown in Fig.~\ref{fig:036037-047-100}(c). The (9-6-3) and (10-6-3) structures shown in Figs.~\ref{fig:knownDTSPs}(e) and \ref{fig:knownDTSPs}(f) have the same structural features. One large sphere is placed at the center of each surface of the unit cell, and the large spheres constituting octahedrons enclose small spheres.

The structural frameworks of the the (10-2-1)$_1$, (10-2-1)$_2$, (10-2-2) structures shown in Figs.~\ref{fig:029-040-100}(b), \ref{fig:029-045046-100}(c), and \ref{fig:038-052-100}(b), respectively are constituted by clathrates of small and medium spheres that enclose one large sphere. The (7-2-1), (8-6-2)$_1$, (8-6-2)$_2$, and (10-4-2) structures, shown in Figs.~\ref{fig:034-044-100}(c), \ref{fig:029-039-100}(b), \ref{fig:035-048-100}(b), and \ref{fig:031-047-100}(b), respectively, have the same structural features with the (13-3-1) and (16-2-2) structures which have already been discovered in Ref.~\cite{PhysRevE.104.024101}, shown in Figs.~\ref{fig:knownDTSPs}(h) and \ref{fig:knownDTSPs}(i).

For some DTSPs, structural frameworks are comprised of small spheres. In the (13-2-1) structure shown in Fig.~\ref{fig:knownDTSPs}(g), one large sphere is surrounded by 24 small spheres consisting of the truncated octahedron, where large spheres constitute the FCC structure without contact. Besides, one medium sphere in a tetrahedral site is also surrounded by 12 small spheres consisting of the truncated tetrahedron, and additionally, one small sphere in an octahedral site is surrounded by 12 small spheres consisting of the cuboctahedron. Since the radii of small and medium spheres are too large as listed in Table \ref{table:radius_ratios_of_13-2-1} to be placed in tetrahedral and octahedral sites, the structural framework can be regarded as one of the unique small-sphere frameworks. 

As the radii of small and medium spheres are getting small, structural frameworks tend to be constituted by large spheres; small and medium spheres are placed in voids among large spheres. For example, in the $\mathrm{XYZ}_4$ structure, which had already been discovered in Ref.~\cite{PhysRevE.104.024101}, large spheres constitute the FCC structures with contact, and small and medium spheres are placed in the tetrahedral and octahedral sites. The (2-2-2)$_1$ and (2-2-2)$_2$ structures shown in Figs.~\ref{fig:knownDTSPs}(a) and \ref{fig:knownDTSPs}(b) are also composed of large-sphere frameworks; small and medium spheres are placed in the polyhedrons that are composed of large spheres. Besides, in the (5-4-3) and (6-4-3) structures shown in Figs.~\ref{fig:034-044-100}(b) and \ref{fig:033-044045-100}(c), respectively, clusters consisting of small and medium spheres are placed in voids among large spheres.

However, some DTSPs might be better not to be classified as those. For example, in the (4-2-2)$\mathrm{_2}$ and (4-2-2)$\mathrm{_3}$ structures shown in Figs.~\ref{fig:029-063064-100}(c) and \ref{fig:029-065-100}(b), respectively, small and medium spheres are placed in voids comprised by large spheres, but simultaneously they loosely enclose the large spheres, as it were to say structural frameworks are constituted by three kinds of spheres. Besides, some DTSPs can also be understood as the cluster-type DTSPs, for example, the (13-3-1) structure shown in Figs.~\ref{fig:knownDTSPs}(h) contains clusters constituted by 13 small spheres. This structure can be derived from a perovskite structure; one site is occupied by the cluster. Furthermore, as discussed in Ref.~\cite{PhysRevE.104.024101}, some of DTSPs such as the (4-3-1) and (4-4-2) structures shown in Figs.~\ref{fig:knownDTSPs}(c) and \ref{fig:knownDTSPs}(d), respectively, can be derived from the DBSPs.

The random approaches can find any type of structure without a priori knowledge, but they are difficult to explore the complex structures consisting of large unit cells. An improved methodology to generate more promising initial structures based on the structural properties of the DTSPs are essential to explore the DTSPs composed of large unit cells.

\subsection{DTSPs and Crystal Structures}
\label{sec:dtsps_and_crystal_structures}

In recent years, many algorithms are developed to predict crystal structures~\cite{GLASS2006713, doi:10.1063/1.2210932, Oganov_2008, LYAKHOV20101623, LYAKHOV20131172, 10.1039/9781788010122, PhysRevB.82.094116, WANG20122063, Wang_2015, WANG2016406, Pickard_2011}. Those methods have successfully predicted many materials, but it is still difficult to find stable structures of ternary and quaternary materials since the size of the configuration space becomes explosively large. Then, we expect that the DTSPs give the unknown structural prototypes of materials under high pressure.

As discussed in Ref.~\cite{PhysRevE.103.023307}, many crystals can be understood as DBSPs. The result implies that the DSPs can be effectively used as structural prototypes for crystal structures, especially for high-pressure phases. In this study, we also investigate the correspondence of the discovered DTSPs with crystals registered in ICSD~\cite{ICSD}. As a result, we have discovered three correspondences; first, we confirm the correspondence between the (4-2-2)$_5$ structure and the ternary $\mathrm{Cu}_3 \mathrm{Ti}$ structure, but this correspondence have already been discussed in the previous study~\cite{PhysRevE.104.024101} as one of the SDTSPs; second, the (4-2-2)$_6$ structure corresponds to the $\mathrm{Cu}_2 \mathrm{GaSr}$ structure while this correspondence was also discussed in the previous study~\cite{PhysRevE.104.024101} as one of the SDTSPs; finally, the (2-1-1) structure~\cite{PhysRevE.104.024101}, which is equal to the (4-2-2)$\mathrm{_6}$ structure with the correction of structural distortion, also corresponds to the $\mathrm{Cu}_2 \mathrm{GaSr}$ structure. Note that we referred to the space group determined by the code Spglib~\cite{togo2018textttspglib} with correcting the structural distortions in each DTSP so that we determine the correspondence. 

The correspondences of the DTSPs and the crystals seem to be exceptional, but some materials comprised of the DTSPs may be realized in future experiments. Possibly, the quaternary and/or quinary structural prototypes, in which some spheres in the DTSPs are replaced by fourth and/or fifth spheres, might be favorable structural prototypes for materials under high pressure. The (13-3-1) structure includes the cluster structures consisting of 13 small spheres, as shown in Fig.~\ref{fig:knownDTSPs}(h), but one small sphere in the cluster is better for nature to be replaced by fourth sphere, where the cluster structure is similar to that constituted by 13 hydrogen atoms in the $\mathrm{MgH}_{13}$ under pressure, which possesses high predicted $T_c$~\cite{PhysRevB.104.054501}. Besides, the (9-6-3) structure shown in Fig.~\ref{fig:knownDTSPs}(e) has one small sphere at the center of the unit cell, but nature might prefer the (8-1-6-3) structure in which a small sphere at the center is replaced by a fourth sphere as shown in Fig.~\ref{fig:prototypes}(a). Boron atoms sometimes constitute the octahedron as can be seen in in the $\mathrm{UB}_4$, so the medium spheres in the (8-1-6-3) structure might be realized by boron atoms. Similarly, the (8-1-6-1-3) structure shown in \ref{fig:prototypes}(b) might be more preferable than the (9-7-3) structure shown in Fig.~\ref{fig:036037-047-100}(c). The (8-1-6-1-3) structure has a fourth sphere at the center of the unit cell replacing a small sphere, and a fifth sphere at the vertex of the unit cell replacing the medium sphere. The fourth and fifth spheres are surrounded by six spheres constituting octahderons, respectively. Finally, the (14-1-1-3) structure, in which all the medium spheres in the (8-1-6-1-3) structure are replaced by small spheres, might be one of the promising structural prototypes for hydrides.

Otherwise, the local structures in the DTSPs might be useful for the design of unknown crystal prototypes. In fact, the (10-4-1) structure shown in Fig.~\ref{fig:10-4-1} includes the tetrahedron consisting of small spheres in the tetrahedral sites comprised by large spheres, and the $\mathrm{LaBeH}_8$~\cite{PhysRevLett.128.047001} also includes the tetrahedron consisting of hydrogen atoms in the tetrahedral sites. The similarity indicates that the dense local structures are preferred by materials under pressure. Therefore, local structures that sometimes appear in the DSPs may be building blocks in crystals. For example, the octahedrons consisting of eight spheres might be preferred by materials under pressure since the octahedrons sometimes appear in the DSPs such as the (16-4)~\cite{PhysRevE.103.023307}, (6-6-4), and (9-6-3) structures.

\section{conclusion}
\label{sec:conclusion}

We exhaustively explored DTSPs at 451 kinds of radius ratios and 436 kinds of compositions. As a result, we have additionally found 22 putative DTSPs, and besides, we successfully confirmed that some of the well-ordered DTSPs that have already been discovered in the previous study~\cite{PhysRevE.104.024101} appear on the ternary phase diagrams at relatively many radius ratios. The 60 putative DTSPs have been identified in total in the previous~\cite{PhysRevE.104.024101} and present studies. Our results show that at a considerable number of radius ratios, the highest packing fraction at each composition is achieved by the phase separations consisting of only the FCC and/or the DBSPs, and the tendency is getting evident as the small and medium spheres are getting larger. This result indicates that the densely-packed local structures composed of three kinds of spheres might be getting more complex as the small and medium spheres are getting larger, and accordingly, the unit cells of the DTSPs composed of three kinds of spheres of similar size are getting larger. Note that this result might directly indicate the local structures in the DBSPs may be denser than any other local structures composed of three kinds of spheres, and accordingly, the DBSPs may be the densest.

Since the minimum radius of small spheres is set to be larger than in the previous study~\cite{PhysRevE.104.024101}, the discovered DTSPs are well-ordered and they tend to have high symmetries. We classify the discovered DTSPs based on how structural frameworks are comprised, and the properties might be applicable to the generation of initial structures. Since our random approach, which just generates multilayered initial structures, seems to be difficult to find the DTSPs composed of larger unit cells than in this study, it may be necessary to develop more effective approaches to create more promising initial structures, for example, we had better generate initial structures in which the small and medium spheres loosely enclose large spheres.

Despite the great progress of the algorithms to predict crystal structures~\cite{GLASS2006713, doi:10.1063/1.2210932, Oganov_2008, LYAKHOV20101623, LYAKHOV20131172, 10.1039/9781788010122, PhysRevB.82.094116, WANG20122063, Wang_2015, WANG2016406, Pickard_2011}, it is still difficult to search the stable structures of ternary and quaternary materials. Then, we expect that the DTSPs may be effectively utilized as structural prototypes to explore the ternary materials under high pressure. Possiblely, the quaternary and/or quinary structural prototypes, in which some spheres in the DTSPs are replaced by fourth and/or fifth spheres, might be favorable structural prototypes for materials under high pressure. The (13-3-1) structure includes the cluster structures consisting of 13 small spheres, as shown in Fig.~\ref{fig:knownDTSPs}(h), but one small sphere in the cluster is better for nature to be replaced by fourth sphere, where the cluster is similar to that of the hydrogen atoms in $\mathrm{MgH}_{13}$ under pressure, which possesses high predicted $T_c$~\cite{PhysRevB.104.054501}.

Our exhaustive exploration of DTSPs reveals that packings of multi-sized hard spheres yield diverse 
periodic structures realized by subtle combinations with a large number of spheres, and indicates that 
further studies will disclose highly ordered structures in larger unit cells including many spheres 
more than those surveyed in this study. The explorations for the densest structures of the quaternary 
sphere packings and these higher dimensional cases will be interesting future directions.

\begin{acknowledgments}

R. K. is financially supported by the Quantum Science and Technology Fellowship Program (Q-STEP), the University Fellowship 
program for Science and Technology Innovations.
The calculations was performed by the supercomputer, Ohtaka, in ISSP, the Univ. of Tokyo.

\end{acknowledgments}

\bibliographystyle{apsrev4-2}
\bibliography{submit}

\end{document}